\documentclass{aa}
\usepackage{graphicx}
\usepackage{psfig}
\usepackage{natbib}
\bibpunct{(}{)}{;}{a}{}{,}

\begin{document}

\def\spose#1{\hbox to 0pt{#1\hss}}
\def\lta{\mathrel{\spose{\lower 3pt\hbox{$\mathchar"218$}}
     \raise 2.0pt\hbox{$\mathchar"13C$}}}
\def\gta{\mathrel{\spose{\lower 3pt\hbox{$\mathchar"218$}}
     \raise 2.0pt\hbox{$\mathchar"13E$}}}
\def\Msun{{\rm M}_\odot}
\def\msun{{\rm M}_\odot}
\def\Rsun{{\rm R}_\odot}
\def\Lsun{{\rm L}_\odot}
\def\half{{1\over2}}
\def\RL{R_{\rm L}}
\def\zs{\zeta_{s}}
\def\zR{\zeta_{\rm R}}
\def\dJJ{{\dot J\over J}}
\def\dMM{{\dot M_2\over M_2}}
\def\tKH{t_{\rm KH}}
\def\eck#1{\left\lbrack #1 \right\rbrack}
\def\rund#1{\left( #1 \right)}
\def\wave#1{\left\lbrace #1 \right\rbrace}
\def\dd{{\rm d}}
\def\new#1{{#1}}
\def\new2#1{{#1}}

\title{The evolution of the timing properties of the black-hole transient
         GX 339--4 during its 2002/2003 outburst}

\titlerunning{Timing GX 339--4 during its 2002/2003 outburst}
\authorrunning{T. Belloni et al.}

\author{T. Belloni\inst{1}, 
                J. Homan\inst{1,2}, 
                P. Casella\inst{1,3,4},
                M. van der Klis\inst{5}, 
                E. Nespoli\inst{1},
                W.H.G. Lewin\inst{2}, 
                J.M. Miller \inst{6}
                \and M. M\'endez\inst{7}
}

\offprints{T. Belloni (belloni@merate.mi.astro.it)}

\institute{INAF -- Osservatorio Astronomico di Brera,
        Via E. Bianchi 46, I-23807 Merate, Italy
   \and
        Center for Space Research, Massachusetts Institute of Technology,
        77 Massachusetts Avenue, Cambridge, MA 02139-4307, USA
   \and
        INAF -- Osservatorio Astronomico di Roma,
        Via di Frascati 33, I-00040 Monte Porzio Catone, Italy
   \and
        Dipartimento di Fisica, Universit\`a degli Studi ``Roma Tre'',
        Via della Vasca Navale 84, I-00146 Roma, Italy
   \and
        Astronomical Institute ``Anton Pannekoek'', University of Amsterdam, 
        and Center for High Energy Astrophysics, Kruislaan 403, 
        NL-1098 SJ, Amsterdam, the Netherlands
   \and
        Harvard-Smithsonian Center for Astrophysics, 
        60 Garden Street, Cambridge, MA 02138, USA
   \and
        SRON, National Institute for Space Research, Sorbonnelaan 2,
        3584 CA Utrecht, the Netherlands 
}

\date{Received 1 December 2004; accepted 15 April 2005}

\abstract{We present the results of the timing and color analysis of
more than two hundred RXTE/PCA observations of the bright black-hole
transient GX 339--4 obtained during its 2002/2003 outburst. The
color-intensity evolution of the system, coupled to the properties of
its fast time variability, allow the identification of four separate
states. Depending on the state, {\new2 strong noise is detected},
together with a variety of quasi-periodic oscillations at
frequencies from 0.2 to 8 Hz. We present a characterization
of the timing parameters of these states and compare them to what has
been observed in other systems. These results, together with those
obtained from energy spectra, point towards a common evolution of
black-hole  transients through their outbursts. 

}

\maketitle

\section{Introduction}

Since the early years of X-ray astronomy, black-hole candidate X-ray
binaries have been known to show transitions between different
spectral states.  When X-ray instrumentation became sufficiently
sophisticated to allow detailed timing studies on short time scales,
the definitions of states were refined to include fast timing
properties (see van der Klis 1995,2005; McClintock \& Remillard
2005). The number and defining properties of these states have changed
with time (see e.g.\ Homan et al. 2001), but it is now clear that fast
timing variations are a key ingredient which needs to be considered in
order to have a complete view of the states and state transitions.

Three states are relatively well identified (see McClintock \&
Remillard 2005; van der Klis 2005).  In the High/Soft State (HS), the
energy spectrum is dominated by a soft thermal component, modeled with
a disk-blackbody (Mitsuda et al. 1984) with a typical temperature of
$\sim$1 keV. This component is attributed to an optically--thick but
geometrically--thin accretion disk (Shakura \& Sunyaev 1976). Very
little short-term variability, in the form of a power-law shaped noise
component, is observed (see e.g. Belloni et al. 1999).  The second is
the Low/Hard State (LS): the energy spectrum is dominated by a hard
component, often associated with a Comptonizing region in the
accretion flow, with a high-energy cutoff energy of $\sim$100 keV.
The power spectra can be decomposed in a small number of flat-top (or
broad) components, approximated with Lorentzian functions, plus at
times a low-frequency quasi-periodic oscillation (QPO; see Nowak 2000;
Psaltis, Belloni \& van der Klis 1999; Belloni, Psaltis, \& van der
Klis 2002; Pottschmidt et al. 2004). The characteristic frequencies of
these components follow rather precise correlations.  Low-luminosity
neutron star X-ray binaries show very similar power spectra (see van
der Klis 1995; Olive et al. 1998; Wijnands \& van der Klis 1999;
Psaltis, Belloni \& van der Klis 1999; Belloni, Psaltis \& van der
Klis 2002; van Straaten et al.  2002,2003), with frequencies following
the same correlation, indicating a probable common origin.  The third
is the Quiescent state, which appears at low luminosity levels and
characterizes the long periods of quiescence of transient systems. The
timing properties here are not well known, given the low flux,
although observations of XTE J1650--500 at a luminosity level of $\sim
10^{34}$erg/s still suggest the presence of Low/Hard State timing
properties (Tomsick, Kalemci \& Kaaret 2004).

In addition to these states, other behavior is observed in transient
systems, often for much shorter periods.  Their properties are
complex and a simple classification was not possible until now. It is
important to note that these additional states are not only
associated to the transitions between LS and HS (see Homan et al. 2001; 
Belloni 2004; McClintock \& Remillard 2004; Casella et al. 2004).  
These states feature the largest variety of
spectral and timing features. In the 1990's, a state associated to the
highest luminosity observed in two systems, GS 1124--683 and GX 339--4
was called Very High State (VHS; Miyamoto et al. 1991,1993,1994). In
this state, strong low-frequency  (1-10 Hz) QPOs were observed
together with noise variability showing rapid transitions between
band-limited and a strong power-law power-spectral shape, while the
energy spectrum is composed of both a hard and a soft component,
with varying relative contributions. Two different QPO behaviors were
observed depending on the energy spectrum, with fast transitions
between them (Takizawa et al. 1997; Miyamoto et al. 1991).
Overall, two separate ``flavors" of VHS were reported: a hard
one, dominated by the hard component in the energy spectrum and
showing band-limited noise in the power spectrum, and a soft
one spectrally dominated by the soft component and with power-law
noise in the power spectrum.

Later, some VHS properties were also observed at much lower
luminosities and a new state was proposed, called the Intermediate State
(IMS: M\'endez \& van der Klis 1997; Belloni et al. 1997). Homan et
al.  (2001) showed that in XTE J1550--564 VHS properties were found at
several well-separated luminosity levels and they proposed to consider
the VHS as just the highest instance of the IMS.  Both HS and LS were
found to occur over a wide range of luminosities, effectively removing
the need to distinguish between VHS and IMS; Homan et al. (2001)
proposed to consider the VHS as just the highest luminosity instance
of the IMS (see also van der Klis 2005).  Recently, McClintock \&
Remillard (2005) proposed a new scheme for states of black-hole
candidates with a somewhat different naming convention: the VHS was
replaced by a more narrowly-defined Steep Power-Law state (SPL) and by
several types of Intermediate States based on the parameters of the
fitted spectral models. As noted by McClintock \& Remillard (2005),
there is no one-to-one correspondence between these primarily X-ray
spectrally defined states and the time variability states we use here.

Clearly, although there is general agreement about the LS and the HS
as to their definition and their basic properties, the states that
differ from those two are complex and more difficult to classify and
to interpret. Notice that the LS and HS are rather stable states,
which can be observed for months or years, while the remaining states
are often associated to state transitions and show strong and fast
variations in their properties (see e.g.  Miyamoto et al. 1994;
Takizawa et al. 1997; Homan et al. 2001; Nespoli et al. 2003; Casella
et al. 2004). Also, the detection of high-frequency QPOs ($>$20 Hz),
although few are known, seems to be limited to these states (Morgan,
Remillard \& Greiner 1997; Remillard et al. 1999; Cui et al. 2000;
Homan et al. 2001; Miller et al. 2001; Strohmayer 2001a,b; Homan et
al. 2003a; Klein-Wolt, Homan \& van der Klis 2004; Casella et
al. 2004; Homan et al. 2005b). It is also in these states that narrow
and often strong low-frequency ($<$20 Hz) QPOs are observed.

These QPOs have been classified into three separate classes. Wijnands
et al. (1999) and Homan et al. (2001) reported on two different types
of QPOs in the RXTE data of XTE~J1550--564: a broad one (type A), with
a quality factor Q (the QPO frequency divided by the QPO
full-width-half-maximum, FWHM)) of less than 3, and a narrower one
(type B), with a Q larger than 6. Both QPOs were characterized by a
centroid frequency of 6 Hz and associated with a weak red-noise
component, but with different phase-lag behavior. XTE~J1550--564 also
showed the more common QPO-type associated with a flat-top noise
component (see Cui et al. 1999 and Sobczak et al. 2000). Remillard et
al. 2002 dubbed this QPO `type C': its features are a high coherence
(Q $>$ 10), a variable centroid frequency (in the range 0.1 - 10 Hz)
and the simultaneous presence of a strong flat-top noise component
($\sim$10--40\% rms). The type-C QPOs are seen to follow the same
global frequency correlations as those of the low/hard state (Psaltis,
Belloni \& van der Klis 1999, Belloni, Psaltis \& van der Klis
2002). While type-C QPOs are observed in many systems, the other two
are less common. In addition to XTE~J1550--564, type-B QPOs were also
observed in GX~339--4 (Nespoli et al. 2003), GRS~1739--278 (Wijnands
et al. 2001), XTE J1859+226 (Casella et al. 2004), H 1743--322 (Homan
et al. 2005b) and possibly in 4U~1630--47 (Tomsick\& Kaaret 2000),
while type-A QPOs were observed in GX~339--4 (Nespoli et al.  2003),
XTE J1859+226 (Casella et al. 2004), H 1743--322 (Homan et al.
2003b,2005b), and possibly in 4U~1630--47 (Tomsick \& Kaaret 2000;
Dieters et al. 2000) Furthermore, in the light of the A-B-C
classification, the two QPOs observed in GS~1124--683 (Takizawa et al.
1997) and GX 339--4 (Miyamoto et al. 1991) can be tentatively
identified with types B and C, although a detailed re-analysis of
Ginga data is necessary to confirm this association. These
oscillations, whose nature is still not understood, could provide a
direct way to explore the accretion flow around black holes (and
neutron stars). In particular, their association with specific
spectral states and the phenomenology that is emerging indicate that
they are a key ingredient in understanding the physical conditions
that give origin to the different states.

\begin{figure}[h] \resizebox{\hsize}{!}{\includegraphics{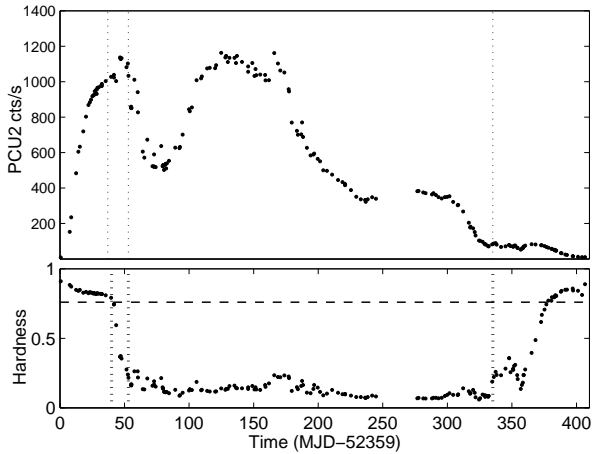}}
\caption{Top panel: PCU2 light curve of the full outburst of GX
339--4, with one point per observation. Bottom panel: corresponding
hardness ratio (see text). The dotted lines indicate major state
transitions (see text). The horizontal dashed line shows the hardness
level corresponding to the transition from the right branch to the left
branch in the HID (see text and Fig. 2).} 
\label{figure1} \end{figure}

State transitions and non LS/HS states appear to be important for the
understanding of accretion onto black holes. There is evidence that
the ejection of relativistic jets takes place during some state
transitions (see Fender, Belloni \& Gallo 2004, Corbel et al. 2004)
and it is during these intervals that structural changes take place in
the accretion flow.

GX 339--4 is a transient black-hole candidate that is known to spend
long periods in outburst. Historically, it was found prevalently in a
hard state, although several transitions were reported (Maejima et
al. 1984; Ilovaisky et al. 1986; Miyamoto et al. 1991). The system was
one of the two that showed all `canonical' X-ray states of BHCs (LS,
HS and VHS: see Miyamoto et al. 1991). From the launch of the Rossi
X-Ray Timing Explorer (RXTE) until 1999 it remained bright, mostly in
the Low/Hard state but with a transition to a softer state (see Nowak,
Wilms \& Dove 1999; Wilms et al. 1999; Belloni et al.  1999; Nowak,
Wilms \& Dove 2002; Corongiu et al. 2003). In 1999, the source went
into quiescence, where it was detected with BeppoSAX at low flux
levels (Kong et al. 2000; Corbel et al. 2003). A new outburst started
in 2002 (Smith et al. 2002a,b; Nespoli et al. 2003; Belloni 2004) and
ended in 2003 (Buxton \& Bailyn 2004).  After roughly one year in
quiescence, a new outburst started, although it has not reached
the same high luminosity levels to date (Buxton et al. 2004; Smith et
al. 2004; Belloni et al. 2004; Homan 2004; Kuulkers et al. 2004;
Israel et al. 2004). A relativistically broadened iron emission line
has been detected in the X-ray spectrum of GX 339--4, indicating the
presence of a non-zero angular momentum in the black hole (Miller et
al. 2004a,b).

Recently, a high mass function (5.8$\pm$0.5 $M_\odot$) has been
measured for the system (Hynes et al. 2003), indicating strong
dynamical  evidence for the black-hole nature of the compact object.
The distance to GX 339--4 is not well known, with a lower limit of
$\sim$6 kpc (see Hynes et al. 2004).

GX 339--4 was the first BHC to show a radio/X-ray correlation in the
LS (Hannikainen et al. 1998; Corbel et al. 2003; Markoff et al.
2003).  Radio observations during the 1999 HS showed clear evidence of
a strong decrease of core radio emission during this state (Fender et
al. 1999b).  During the 2002/2003 outburst, near the transition to the
VHS (see Smith et al. 2002c), a bright radio flare was observed
(Fender et al. 2002), which led to the formation of a large-scale
relativistic jet (Gallo et al. 2004).

It is clear that GX 339--4 is a very important source for our
understanding of the accretion and ejection properties of stellar-mass
black holes. In this paper we present and discuss the results of the
timing and color analysis of 205 RXTE/PCA observations of GX339--4
during the 2002/2003 outburst. Preliminary results were presented in
Belloni (2004). The results of a complete spectral analysis is
presented in a forthcoming paper, while the analysis of the
optical/IR/X-ray correlations is presented in Homan et al.  (2005a).
\new2{Spectral analysis of these data up to 2004 Jan 29 was included
in the long-term study by Zdziarski et al. (2004).}

\begin{figure}[h] \resizebox{\hsize}{!}{\includegraphics{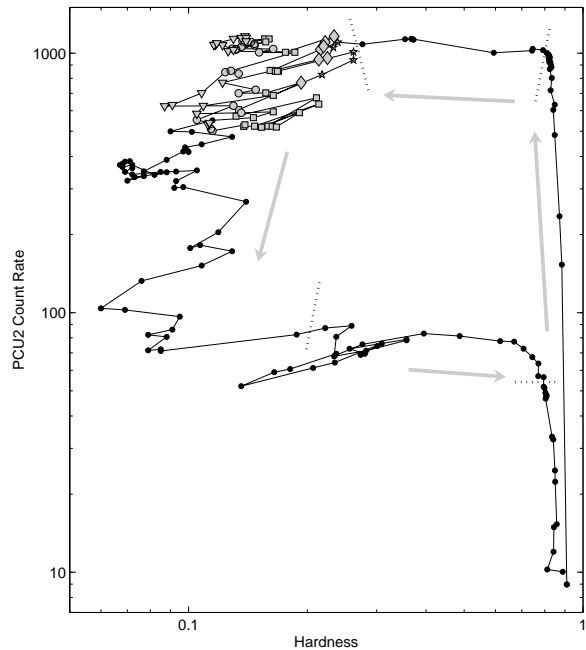}}
\caption{Hardness-Intensity diagram corresponding to the points in
Fig. \ref{figure1}. Each point corresponds to one observation. Dotted
lines correspond to those in Fig. \ref{figure1}. Observation \#0
is indicated by an empty circle. \new2{The gray symbols mark the 
observations included in the groups B (stars), A (diamonds), U1 (squares),
U2 (circles) and weak (triangles) (see section 6). The gray arrows indicate the
general time evolution along the outburst}} \label{figure2} \end{figure}

\section{Data analysis}

We analyzed a large set of 205 RXTE/PCA observations covering the
2002/2003 outburst of GX 339--4, \new2{partly from our own project and
partly from data in the public archive.}  Notice that there are
detections before MJD 52359 (see Homan et al. 2005a), but the source
is too weak to detect signal in the power spectra. The observation log
is shown in Table 1 \new2{(the table is limited to the observations
mentioned explicitly throughout the text. The full table is available
on line)}. The total amount of good data was 555904 seconds. Since the
observations correspond to a few different RXTE proposals, the
observation modes differ between observations.  We produced
background-corrected count rates for each observation in the PCA
channel range 8-49 (3.8-21.2 kev), as well as a hardness ratio defined
as the ratio of counts in the channel range 15-24 (6.3-10.5 keV) over
those in the channel range 8-14 (3.8-6.3 keV).  \new2{As the data span
a long period of time, we checked the stability of the PCA gain by
producing hardness and count rate from two archival observations of
the Crab, from 2002 March 28 and 2003 April 23. Between these two
dates, the Crab count rate was found to increase by 0.02\%, and its
hardness to decrease by 0.2\%. Both values are too small to affect our
results.}

For timing analysis, we used custom timing-analysis software under IDL
and MATLAB. For each observation we produced power spectra from
stretches 128 seconds long using two separate energy bands: PCA
channel band 0-35 (corresponding to 2-15.3 keV) for the main power
spectrum, and PCA channel band 15-40 (6.3-17.3 keV) in order to look
for high-frequency oscillations, which are usually more prominent in
this high-energy band (see Homan et al. 2002).

The power spectra in the high-energy band were also used for a
consistency check on the decomposition of broad-band noise into
Lorentzian components. In the following, for power spectrum we mean
the low-energy power spectrum. For observations with ID 40031 the main
power spectra were produced in the 8-35 (3.8-15.3 keV) channel range
for lack of data at lower energies.  The chosen time resolution was
$1/1024$s, corresponding to a Nyquist frequency of 512 Hz, for the
low-energy data and $1/8192$ for the high-energy data.  Besides
examining the spectrograms, i.e. the time-frequency images, for each
observation we averaged all power spectra together. The high-frequency
($>256$ Hz) part of the average power spectrum (where no signal could
be seen) was fitted with a constant to estimate the level of the
Poissonian noise.  This average was subtracted from the power
spectrum, which was then rebinned logarithmically and converted to
fractional squared rms. The power spectra were then fitted with a
combination of Lorentzians (see Nowak 2000; Belloni, Psaltis \& van
der Klis 2002) using XSPEC v11.3.  \new2{In the following, by flat-top
Lorentzians we mean components whose centroid frequency is much lower
than their width, often consistent with zero.}

We also produced cross-spectra between the data in channel range 0-11
(2-5 keV) and those in the channel range 12-29 (5-12.2 keV) from
stretches of 16-s length. We then calculated averaged cross-spectrum
vectors for each observation, from which we derived phase-lag spectra
(see Casella et al. 2004). Positive phase lags indicate that the hard
curve {\it lags} the soft one. In order to extract the relevant
phase-lag information for different components, we also accumulated
the phase lags in specific frequency ranges: 1-64 Hz to characterize
the full spectrum, 1-6 Hz for the $L_h$ component (see below), and for
the QPO in a range centered on the QPO centroid frequency with a width
equal to the FWHM of the QPO peak itself (see Reig et
al. 2000). Notice that the estimate of the QPO lags depends on the
relative level of the continuum and the QPO powers in the frequency
range considered. When the continuum power is significantly larger
than the QPO power, the lags of the continuum may well dominate.

\begin{table}
\begin{center}
\caption{\small Log of the PCA observations 
\new2{(limited to the ones explicitly
mentioned in the text. The headers mark the four source
states (see text).
The full table is available only in 
electronic form at the CDS.)}}
\begin{tabular}{llcccc}
\hline
\hline
Obs &    Obs. ID  &     MJD             & Exp. & PCU2 & HR\\
\#  &             &                     &   (s)& c/s  &   \\
\hline
\multicolumn{6}{c}{Low-Hard State}\\
\hline
0     & 60705-01-54-00 & 52359.64 & 768   & 9.0    & 0.91 \\
1     & 60705-01-55-00 & 52366.60 & 896   & 153.2  & 0.88 \\
2     & 70109-01-02-00 & 52367.76 & 6784  & 235.3  & 0.87 \\
8     & 70110-01-04-00 & 52381.11 & 1536  & 868.4  & 0.82 \\
17    & 70109-01-04-00 & 52387.56 & 10880 & 956.0  & 0.82 \\
24    & 70110-01-08-00 & 52394.44 & 1536  & 1003.3 & 0.81 \\
\hline
\multicolumn{6}{c}{Hard Intermediate State}\\
\hline
25    & 70110-01-09-00 & 52398.66 & 1024  & 1027.3 & 0.79 \\
27    & 70108-03-01-00 & 52400.85 & 8192  & 1023.4 & 0.74 \\
28    & 70110-01-10-00 & 52402.49 & 1024  & 1003.0 & 0.59 \\
29    & 70109-04-01-00 & 52405.58 & 5888  & 1136.7 & 0.37 \\
30    & 70109-04-01-01 & 52405.71 & 17408 & 1134.3 & 0.37 \\
32    & 70110-01-11-00 & 52406.70 & 896   & 1132.3 & 0.35 \\
33    & 70110-01-12-00 & 52410.53 & 1152  & 1081.7 & 0.28 \\
\hline
\multicolumn{6}{c}{Soft Intermediate State}\\
\hline
34    & 70109-01-07-00 & 52411.60 & 5376  & 1102.5 & 0.24 \\
35    & 70110-01-13-00 & 52412.07 & 896   & 1033.4 & 0.22 \\
41    & 70110-01-14-00 & 52416.60 & 768   & 1011.0 & 0.26 \\
42    & 70110-01-15-00 & 52419.24 & 896   & 940.7  & 0.26 \\
43    & 70108-03-02-00 & 52419.43 & 10368 & 827.2  & 0.22 \\
94    & 70110-01-45-00 & 52524.95 & 1280  & 1161.3 & 0.23 \\
95    & 70110-01-46-00 & 52527.85 & 1024  & 1102.8 & 0.22 \\
96    & 70109-01-23-00 & 52529.58 & 1792  & 1062.7 & 0.22 \\
97    & 70110-01-47-00 & 52532.75 & 1280  & 1051.8 & 0.23 \\
98    & 70110-01-48-00 & 52536.11 & 640   & 957.7  & 0.23 \\
99    & 70109-01-24-00 & 52536.36 & 2816  & 945.1  & 0.21 \\
103b  & 70130-01-02-00 & 52546.51 & 9344  & 769.9  & 0.19 \\
109   & 70110-01-55-00 & 52558.63 & 1024  & 564.4  & 0.15 \\
\hline
\multicolumn{6}{c}{High/Soft State}\\
\hline
110   & 70109-01-26-00 & 52560.41 & 3200  & 550.7  & 0.12 \\
148f  & 70110-01-86-00 & 52693.73 & 896   & 82.3   & 0.19 \\
\hline
\multicolumn{6}{c}{Hard Intermediate State}\\
\hline
149   & 70109-01-37-00 & 52694.92 & 3712  & 87.2   & 0.22 \\
149g  & 50117-01-03-00 & 52706.84 & 7808  & 78.9   & 0.36 \\
149h  & 70110-01-89-00 & 52707.92 & 896   & 72.7   & 0.26 \\
153c  & 50117-01-04-00 & 52715.85 & 17408 & 52.1   & 0.14 \\
154   & 60705-01-57-00 & 52716.70 & 2560  & 59.0   & 0.16 \\
154b  & 70110-01-92-00 & 52717.50 & 1024  & 60.6   & 0.18 \\
155   & 60705-01-59-00 & 52731.56 & 3456  & 77.7   & 0.62 \\
157b  & 70110-01-01-10 & 52740.01 & 896   & 56.4   & 0.79 \\
\hline
\multicolumn{6}{c}{Low/Hard State}\\
\hline
158   & 80116-02-02-00 & 52741.70 & 6016  & 51.8   & 0.79 \\
158b  & 70128-02-03-00 & 52742.24 & 14464 & 49.2   & 0.80 \\
161   & 60705-01-61-00 & 52746.97 & 2816  & 33.3   & 0.84 \\

\hline

\hline
\end{tabular}
\end{center}
\end{table}

\section{The outburst evolution}

\new2{In this section, we describe the general evolution of 
the outburst, as can be followed from Fig. \ref{figure1} and \ref{figure2}.}
In Fig. \ref{figure1} we show the full light curve of the outburst
and the corresponding evolution of the hardness. The
Hardness-Intensity  diagram is shown in Fig. \ref{figure2}. Looking
at Figs. \ref{figure1} and \ref{figure2}, we can identify four
separate sections of the outburst, corresponding to the four sides of
the square traced by the source in the HID. The part of the outburst
covered here starts at the lower right. The source increases its flux
steadily for 36 days, with only a small softening of the spectrum
indicated by the slight slant of the right side of the square. 
Then the source on the HID starts moving rapidly to the left, with the
source changing spectral shape but not its count rate.  Notice that a
constant count rate and a softening of the spectrum indicate that the
source flux must be decreasing.
The transition is indicated in both figures with a dotted line.  The
precise position of this line (between MJD 52394.44 and MJD 52398.66)
is determined by the timing properties (see below), but also by the
properties of the optical/IR/X-ray correlation, which shows a marked 
reversal corresponding to this date (Homan et al. 2005a).
This horizontal branch lasts about 10 days, after which the left
branch is reached (after MJD 52410.53).  The second transition is
also  shown with dotted lines: the precise positions of the
transitions are determined by changes in the timing behavior (see
below). There follows a long period on the left branch, which is
followed  with larger relative variations in hardness than on the right branch.
Also, the count rate is not monotonically decreasing on this branch.
After an initial motion downwards, the source moves up and
reaches a count rate similar to  that of the first peak, near the top
of the right branch (see also Fig. \ref{figure1}), after which it
starts decreasing again. Finally, towards the end of the outburst
(after MJD 52693.73),  there is a monotonic decrease in count rate 
and resulting motion downwards until GX
339--4 enters  the lower horizontal branch, which eventually curves
and becomes once more vertical, ending its outburst on a position
very close to its starting position.  In order to identify these branches
in terms of `canonical' BHC states,  we need to investigate the
timing and spectral properties. A spectral analysis will be presented in 
a forthcoming paper. In the following, we describe the 
\new2{detailed} evolution of
the power spectrum of 
GX 339--4 from a purely phenomenological point of view. The
relation between our results and the source states, as determined
also from spectral parameters, will be examined in the  discussion.

\begin{figure}[h] \resizebox{\hsize}{!}{\includegraphics{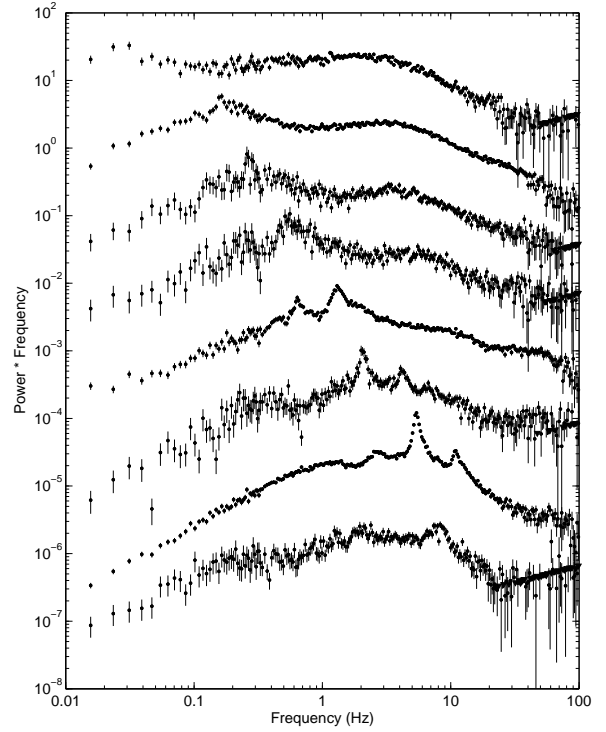}}
\caption{Four average power spectra from the right branch and four from the top 
branch. From top to bottom, they correspond
to observations \#2,17,24,25,27,28,30,33.
 The power spectra are plotted in the
$\nu P_\nu$ representation (Belloni et al. 1997)
and are shifted vertically for clarity. \
The increasing trend at high
frequencies that can be seen in some power spectra is made of 3$\sigma$ upper
limits, shown as inverted triangles.
} \label{figure3}
\end{figure}

\section{The right branch}

The first significant power detected in the power spectrum corresponds to
observation \#0. Here, because of the low count rate, a fit with one
Lorentzian gives a reduced $\chi^2$ close to unity, but the fit is
clearly not good, as considerable excess power is visible at low
frequencies. A fit with two Lorentzians gives as characteristic
frequencies 0.021$\pm$0.7 Hz and 0.32$\pm$0.08 Hz, with a total
integrated fractional rms of $\sim$25\%. As these frequencies are
much lower than the ones observed in subsequent observations,
we do not know how to classify them and do
not include them in the Tables.

Observation \#1
takes place one week after Obs. \#0 and the count
rate is a factor of 16 higher. 
The power spectrum of Obs. \#2 is shown in Fig.
\ref{figure3} (top spectrum). The shape of the power spectrum is very similar to
that observed by RXTE in GX 339--4 during its typical Low/Hard State
(see Belloni et al. 1997; Nowak, Wilms \& Dove 1999)  and also
similar to that 
of most BHCs in that state (McClintock \&
Remillard 2005). From Obs. \#1 (similar to Obs. \#2) to Obs. \#24,
the power spectrum has a similar shape, with its characteristic frequencies
increasing with time and therefore with count rate. All power spectra can be
fitted with up to four Lorentzians: one at low frequencies ($L_1$),
one at intermediate frequencies ($L_2$), one at high frequencies
($L_3$) and a QPO peak ($L_{Q}$, observed only in a few cases). For
the first two observations, an additional Lorentzian component $L_i$
is needed between  $L_1$ and $L_2$. 

\begin{figure}[h] \resizebox{\hsize}{!}{\includegraphics{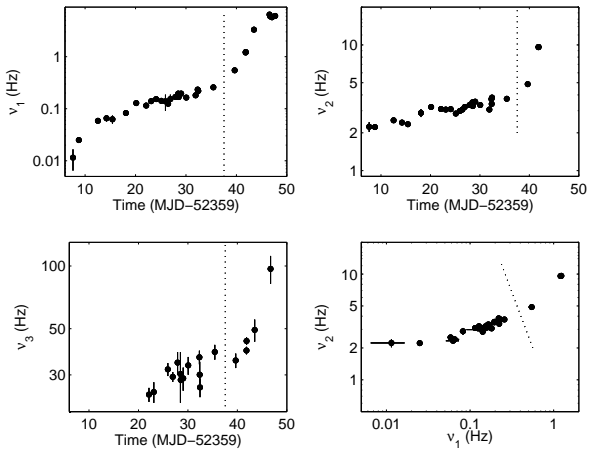}}
\caption{First three panels: time evolution of the characteristic frequencies
of the $L_1$, $L_2$ and $L_3$ components for the \new2{top/right}-branch 
observations at the beginning of the outburst.
Fourth panel: $\nu_2$ vs. $\nu_1$ correlation. In all panels, the dotted
line separates the \new2{right-branch} points from the top-branch points.
} \label{frequencies}
\end{figure}

The characteristic frequencies (defined as $\nu_{max}$,  see Belloni,
Psaltis \& van der Klis 2002) of these components can be seen in
Table 2. Their time evolution, with the exception of the $L_i$
component, present in only two spectra, can be seen in Fig.
\ref{frequencies}, together with the tight correlation between
$\nu_1$ and $\nu_2$.

\begin{table*}
\begin{center}
\caption{\small Main characteristic frequencies (in Hz) for the right and top branches of the HID}
\begin{tabular}{lcccccc}
\hline
\hline
Obs \#& $\nu_{0}$ &$\nu_1$          & $\nu_{Q}$     &  $\nu_2$      &   $\nu_3$      & $\nu_i$      \\
\hline
\multicolumn{7}{c}{Right branch}\\
\hline
1& ...            &0.012 $\pm$0.005 &  ...          & 2.23$\pm$0.20 & ...            & 0.28$\pm$0.06\\
2& ...            &0.03  $\pm$0.01  &  ...          & 2.22$\pm$0.05 & ...            & 0.38$\pm$0.04\\
3& ...            &0.06  $\pm$0.01  &  ...          & 2.52$\pm$0.07 & ...            &     ...      \\
4& ...            &0.07  $\pm$0.01  &  ...          & 2.41$\pm$0.07 & ...            &     ...      \\
5& ...            &0.06  $\pm$0.01  &  ...          & 2.34$\pm$0.08 & ...            &     ...      \\
6& ...            &0.08  $\pm$0.01  &  ...          & 2.88$\pm$0.21 & ...            &     ...      \\
7& ...            &0.13  $\pm$0.01  &  ...          & 3.21$\pm$0.05 & ...            &     ...      \\
8& ...            &0.12  $\pm$0.01  &  ...          & 3.09$\pm$0.07 & 24.14$\pm$1.85 &     ...      \\
9& ...            &0.14  $\pm$0.01  &  ...          & 3.06$\pm$0.09 & 24.85$\pm$2.82 &     ...      \\
10& ...           &0.15  $\pm$0.01  &  ...          & 3.08$\pm$0.08 & ...            &     ...      \\
11& ...           &0.14  $\pm$0.01  &  ...          & 2.85$\pm$0.05 & ...            &     ...      \\
12& ...           &0.14  $\pm$0.05  &  ...          & 2.99$\pm$0.04 & 31.92$\pm$2.28 &     ...      \\
13& ...           &0.13  $\pm$0.01  &  ...          & 3.06$\pm$0.16 & ...            &     ...      \\
14& ...           &0.15  $\pm$0.03  &  ...          & 3.21$\pm$0.03 & 29.38$\pm$1.61 &     ...      \\
15& ...           &0.17  $\pm$0.01  &  ...          & 3.33$\pm$0.05 & 34.40$\pm$4.23 &     ...      \\
16& ...           &0.20  $\pm$0.02  &  ...          & 3.49$\pm$0.12 & 30.32$\pm$8.30 &     ...      \\
17& ...           &0.17  $\pm$0.01  & 0.16$\pm$0.01 & 3.26$\pm$0.03 & 28.38$\pm$0.97 &     ...      \\
18& ...           &0.20  $\pm$0.01  & 0.20$\pm$0.01 & 3.54$\pm$0.08 & 28.87$\pm$3.66 &     ...      \\
19& ...           &0.16  $\pm$0.01  &  ...          & 3.33$\pm$0.08 & 33.34$\pm$3.26 &     ...      \\
20& ...           &0.18  $\pm$0.01  &  ...          & 3.06$\pm$0.11 & ...            &     ...      \\
21& ...           &0.23  $\pm$0.01  & 0.20$\pm$0.01 & 3.71$\pm$0.07 & 36.48$\pm$2.74 &     ...      \\
22& ...           &0.22  $\pm$0.01  &  ...          & 3.39$\pm$0.10 & 30.08$\pm$4.60 &     ...      \\
23& ...           &0.22  $\pm$0.01  &  ...          & 3.81$\pm$0.14 & 26.15$\pm$2.68 &     ...      \\
24& ...           &0.26  $\pm$0.02  &  ...          & 3.72$\pm$0.08 & 38.68$\pm$3.14 &     ...      \\
25& ...           &0.55  $\pm$0.05  &  ...          & 4.88$\pm$0.16 & 35.20$\pm$2.81 &     ...      \\
\hline
\multicolumn{6}{c}{Top branch}\\
\hline
26& ...           &1.21  $\pm$0.03  & 1.26$\pm$0.01 & 9.59$\pm$0.23 & 39.31$\pm$1.64 &    ...       \\
27& ...           &1.23  $\pm$0.02  & 1.33$\pm$0.01 & 9.63$\pm$0.31 & 43.65$\pm$1.85 &    ...       \\
28& ...           &3.31  $\pm$0.12  & 2.08$\pm$0.01 & ...           & 49.33$\pm$6.03 &    ...       \\
29& 1.08$\pm$0.01 &6.46  $\pm$0.26  & 5.46$\pm$0.01 & ...           &        ...     &    ...       \\
30& 0.99$\pm$0.01 &5.93  $\pm$0.13  & 5.45$\pm$0.01 & ...           & 96.93$\pm$14.78&    ...       \\
31& 0.94$\pm$0.02 &5.73  $\pm$0.61  & 5.34$\pm$0.02 & ...           &        ...     &    ...       \\
32& 0.99$\pm$0.01 &6.08  $\pm$0.75  & 5.82$\pm$0.02 & ...           &        ...     &    ...       \\
33& 0.99$\pm$0.01 & ...             & 8.12$\pm$0.18 & ...           &        ...     &    ...       \\

\hline
\end{tabular}
\end{center}
\end{table*}

The integrated fractional rms of the $L_1$ component decreases
smoothly from about 30\% to $\sim$17\%, while at the same time the
$L_2$ component goes from 28\% to 14\%. The fractional rms of the
$L_3$ component, when present, is always 2-3\%, and the few
detections of  $L_{Q}$ are at a level of 4-7\%.
The quality factor $Q$ 
is always less than unity for $L_1$ and
$L_2$, $\sim$1 for $L_3$ and typically $\sim$8 for $L_Q$.

\begin{figure}[h] \resizebox{\hsize}{!}{\includegraphics{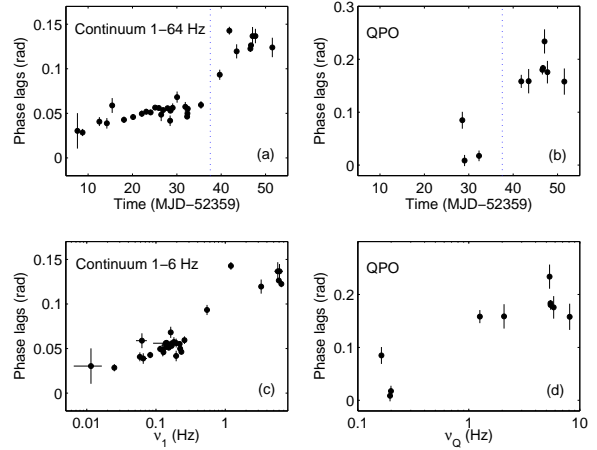}}
\caption{Phase lags for the \new2{right} and top branches. (a) Time evolution of
        the lags of the 1-64 Hz continuum. (b) Time evolution of the
        QPO lags. (c) Correlation between the lags of the 1-6 Hz continuum
        and the characteristic frequency $\nu_1$ of the $L_1$ component.
        (d) Correlation between the lags of the QPO and its characteristic
        frequency.
} \label{lags}
\end{figure}

In order to identify the broad components with those presented by
Belloni, Psaltis \& van der Klis (2002), we need more detections of
the QPO component. Therefore, this identification will be shown at the
end of the next section. In any case, all the power spectra
corresponding to this branch of the HID have a shape that allows us to
classify these observations as belonging to the low/hard state.

The phase lags for the continuum, for the QPO, and between 1 and 6 Hz
can be seen in Fig. \ref{lags}. All lags are positive.  For the QPO
there are only three points, but the other two show a clear  increase
in the value of the hard lag from less than 0.05 radians at the start
to almost 0.15 radians at the top of the right branch.  A
representative phase-lag spectrum (from Obs. \#8) is shown in Fig.
\ref{lagspectra}. Hard lags are also typical of the low/hard state 
(see e.g Pottschmidt et al. 2000).

\section{The top branch}

\begin{figure}[h] \resizebox{\hsize}{!}{\includegraphics{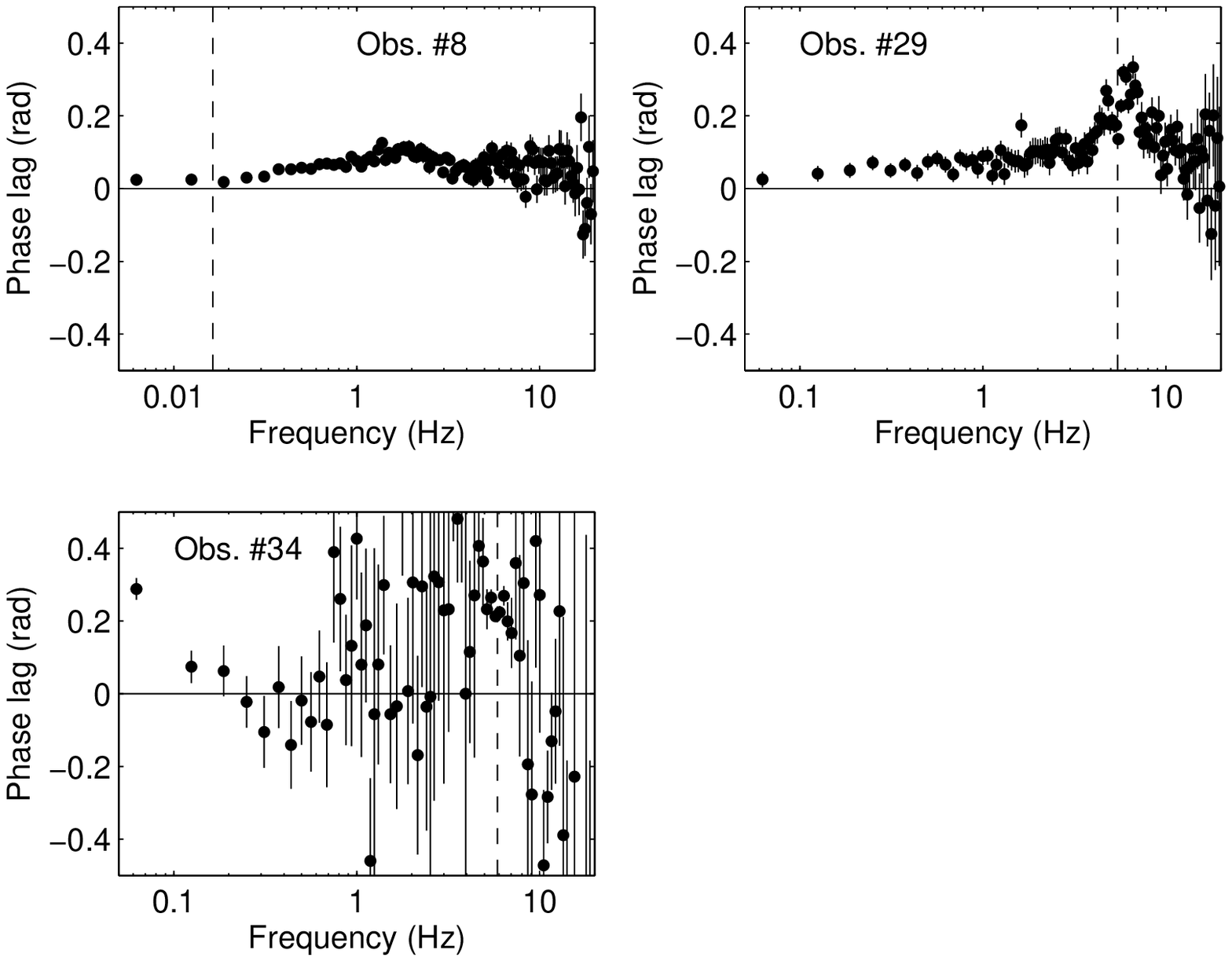}}
\caption{Phase-lag spectra for three representative observations from 
        the right (Obs. \#8), top (Obs. \#29) and left (Obs. \#34) branches.
        The vertical dashed lines mark the centroid frequency of the QPO.
} \label{lagspectra} 
\end{figure}

\begin{figure}[h] \resizebox{\hsize}{!}{\includegraphics{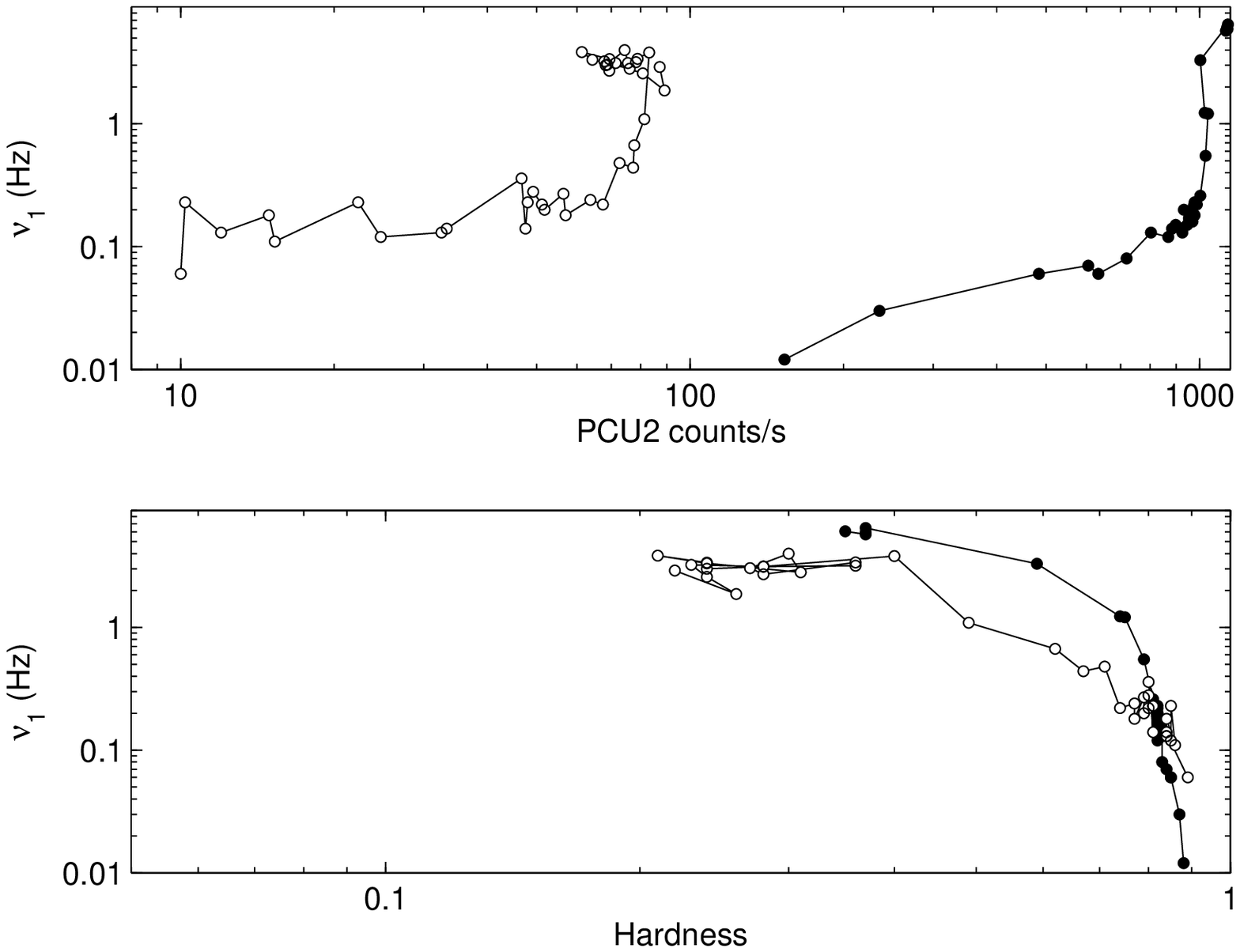}}
\caption{Evolution of the $\nu_1$ frequency as a function of count rate
        (top panel) and hardness (bottom panel) for the observations of the 
        right and top branches (filled circles) and of the bottom branch
        (empty circles). The X axes of both panels are the same as on 
        the corresponding axis on Fig. \ref{figure1}} 
        \label{nu_rate_color}
\end{figure}

Starting from Obs. \#25, the hardness ratio starts to decrease more
rapidly (see Fig.  \ref{figure2}) and a clear QPO with harmonic
content appears in the power spectrum. This can be seen in the bottom
four spectra from Fig. \ref{figure3}, which also shows that the
characteristic frequencies continue to increase. The shape of the
power spectrum is typical of the Very High State (VHS) in its hard
version with band-limited noise (see Miyamoto et al. 1991,1993,1994;
Homan \& Belloni 2005).

In order to fit these power spectra, additional components such as
harmonic peaks and QPO shoulders (see e.g. Belloni et al. 1997) are
needed. We decided not to include the shoulders in our fits in order
to use a model consistent with that used for the previous
observations. The only additional component is a broad Lorentzian
($L_0$) appearing at low frequencies, while the $L_3$ component
becomes undetectable, likely because its characteristic frequency
increases and leaves the frequency range where we detected significant
power.  The characteristic frequencies are reported in Table 2 and
their evolution is shown in Fig. \ref{frequencies}.  It is evident
that the components detected in the top branch are consistent with
being the extension to high frequency of those seen in the right
branch.  However, a \new2{gradual} transition can be seen in the top
right panel of Fig. \ref{frequencies}.  Despite the fact that the
energy spectrum changes considerably, as indicated by the hardness
ratio, the timing properties do not seem to show sharp changes, with
the exception of the appearance of a clear QPO of type C (see Miyamoto
et al. 1994, Remillard et al.  2002 and Casella et al. 2004). The
characteristic frequencies increase, and in particular $\nu_1$ and
$\nu_2$ show a fast time evolution to higher values. The fourth panel
in Fig. \ref{frequencies} shows their correlation.
\new2{Figure \ref{nu_rate_color} shows the evolution of $\nu_1$ for
the right and top branches (filled circles) as a function
of count rate and hardness. The gradual transition between the two branches is 
evident.}

\begin{figure}[h] \resizebox{\hsize}{!}{\includegraphics{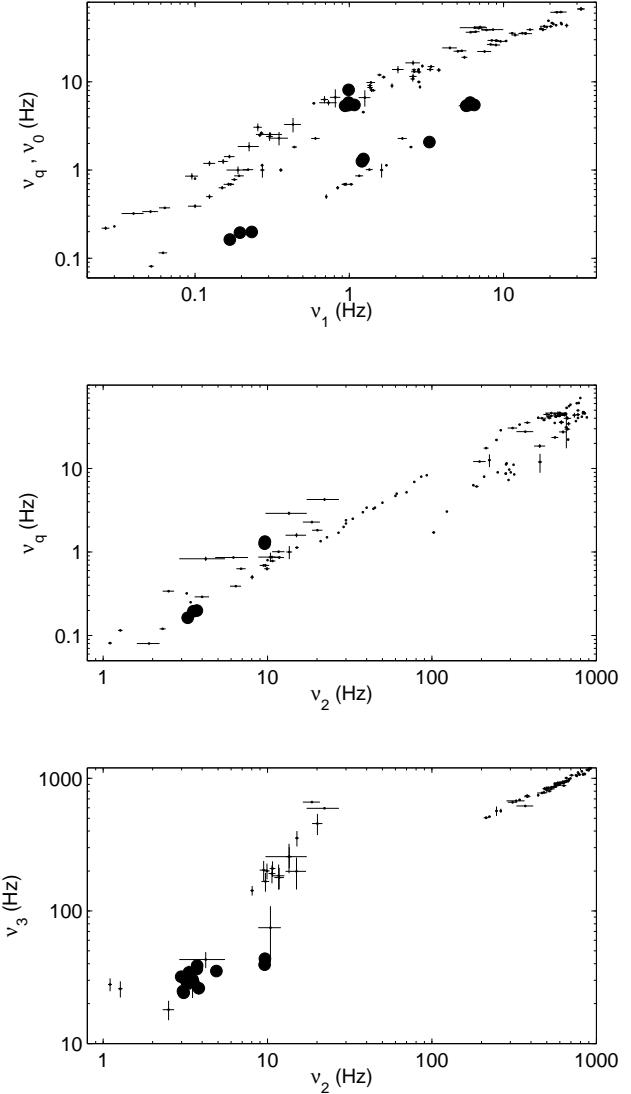}}
\caption{Top panel: correlation between $\nu_1$ and $\nu_Q$ (large circles) and
        points from the WK correlation (Wijnands \& van der Klis 1999, Belloni,
        Psaltis \& van der Klis 2002). The upper points are our detections
        of $\nu_0$ from the top branch.
        Middle panel: correlation between $\nu_Q$ and $\nu_2$ (large circles) 
        and points from the PBK 
        correlation (Psaltis, Belloni \& van der Klis
        1999, Belloni, Psaltis \& van der Klis 2002).
        Bottom panel: correlation between $\nu_3$ and $\nu_2$ (large circles) 
        and points from the PBK second correlation 
        (Psaltis, Belloni \& van der Klis
        1999, Belloni, Psaltis \& van der Klis 2002).
} \label{correlations} 
\end{figure}

We can now attempt an identification of the components present in our
power spectra and those presented in Wijnands \& van der Klis (1999),
Psaltis, Belloni \& van der Klis (1999) and Belloni, Psaltis \& van
der Klis (2002). It is natural to identify the $L_Q$ component with
the low-frequency QPO in BHC. The characteristic frequency of the
$L_1$ component is always close to $\nu_Q$ and on that basis can be
identified with the $L_h$ component from Belloni, Psaltis \& van der
Klis (2002). Following this scheme, $L_0$ corresponds to $L_b$, L$_2$
to $L_\ell$ and $L_3$ to $L_u$ (see Belloni, Psaltis \& van der Klis
2002 for the description of these components). In order to check
whether these identifications hold, we plot the three global
correlations (WK from Wijnands \& van der Klis 1999; and PBK from
Psaltis, Belloni \& van der Klis 1999) in Fig. \ref{correlations},
together with the corresponding points from our observations.  It is
worth noting that the $L_b$ ($L_0$) component is present here in only
a few observations, while it constitutes the dominant component in
other systems (see Wijnands \& van der Klis 1999). This is also seen
in many power spectra of GRS 1915+105 (see e.g. Morgan, Remillard \&
Greiner 1997).

Due to this, we can put only four points on the WK relation. Also
shown in Fig. \ref{correlations} (top panel) are the $\nu_1 - \nu_Q$
pairs, which all have $\nu_1\approx \nu_Q$. This second correlation
was not present in the original Wijnands \& van der Klis (1999) work,
but was presented by Belloni, Psaltis \& van der Klis (2002).

For comparison, from the early-outburst observations of XTE
J1550--564, one can see that the $L_b$ and $L_h$ components have a
rather different energy spectrum, with $L_h$ appearing much more
prominent at higher energies (see Fig. 4 in Cui et al. 1999). However,
a check from the high-energy power spectrum from our observations
shows that $L_0$ does not appear at higher energies.
Fig. \ref{correlations} suggests that the identification $\nu_1 =
\nu_h$ and $\nu_2 = \nu_\ell$ is correct: in particular, the
alternative interpretation $\nu_1 = \nu_b$ and $\nu_2 = \nu_h$
provides no match at all to the WK relation in the top panel of
Fig. \ref{correlations}.  The agreement visible in
Fig. \ref{correlations} shows that the identification is correct.

It is interesting to notice that the integrated fractional rms of the
$L_1$ component continues its smooth decrease observed in the LS,
reaching $\sim$8\% at Obs. \#32, while the rms of $L_2$ drops to 
a few \% before the component becomes undetectable. The QPO has
a rather stable rms between 4-5\%, similar to that of the few
detections along the right branch.

The evolution of the phase lags (see previous section) along the top
branch, after the transition, can be seen in Fig. \ref{lags}, while one
representative phase-lag spectrum (for Obs. \#29) is shown in 
Fig. \ref{lagspectra}. Once again, hard lags are observed. Also the
lags seem to show a smooth evolution through the state-transition.
These phase lags are in
marked contrast with what is usually observed in VHS observations, where
the type-C QPO shows {\it soft} lags (see Wijnands et al. 1999;
Reig et al. 2000;
Homan et al. 2001; Remillard et al. 2002; Casella et al.  2004).

The last observation of this state, Obs. \#33, four days after
Obs. \#32, shows a higher QPO frequency
($\sim$7.8 Hz) which can be classified as type C$^*$ (see Casella 
et al. 2004). Only the $L_0$ component can be detected here. 
An additional band-limited component appears at lower
frequencies, with a characteristic frequency $\sim$0.16 Hz.

\begin{figure}[h] \resizebox{\hsize}{!}{\includegraphics{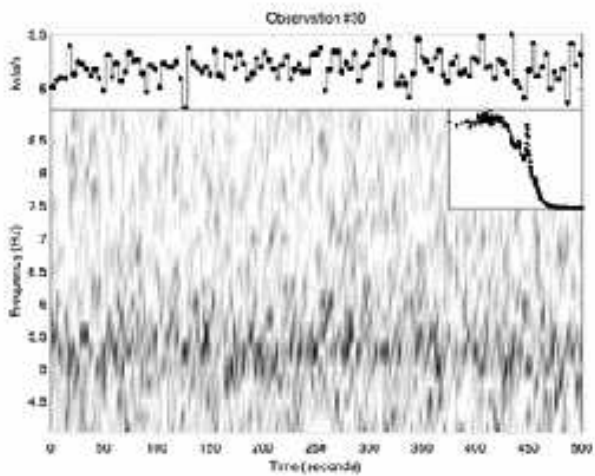}}
\caption{Top panel: a 500-s segment of the light curve of Obs. \#30 (1 second
bin size); count rate is total count rate for all PCUs. 
Bottom panel: corresponding spectrogram (time integration 4 seconds,
time step 1 second), where darker points correspond to higher power. Inset:
average power spectrum from the full observation.
} \label{b5}
\end{figure}

In order to compare type-C QPOs with type-B QPOs, in Fig. \ref{b5},
we show the spectrogram for Obs \#30 (see Sect. 6.1). No clear time
variability of the centroid frequency of the type-C QPO is seen.

\section{The left branch (top section)}

After Obs. \#33, the power spectrum 
changes radically and no more type-C QPOs are
observed until the source leaves the left branch on the HID. As one
can see from Fig. \ref{figure2}, after having crossed the second dotted line,
GX 339--4 spends a long time to its left. During this period, the flux 
does not change monotonically (see Fig. \ref{figure1}).
This complex interval continues up to observation \#109 (MJD 52558). After
this time, the PCU2 count rate remains below 500 cts/s and 
only a weak power-law signal is observed in the power spectrum: this 
bottom section of the left branch will be described in the next section.
During the period before Obs. \#109, 
the timing behavior of GX 339--4 is complex. We summed 
the power spectra of consecutive observations which were close in time and had
a similar position in the HID. The resulting power spectra can be summarized in 
a few types (see Table 3):

\begin{itemize}

\item {\it no signal}: \new2{no significant variability is detected in the 
        power spectrum.}

\item {\it red noise}: a featureless red-noise component can be seen.

\item {\it weak QPO}:
        typically, in the low-energy band, a broad QPO is detected
        at a frequency of 
        $\sim$5 Hz, together with a broad-band noise component with a
        characteristic frequency of a few Hz.
        In the high-energy band a 
        QPO at $\sim$10 Hz is also seen. All these features have an integrated
        fractional rms of a few \%.

\item {\it low-frequency noise}: some low-frequency power in the form of
        a weak band-limited noise is observed, with no significant QPO
        detection either at low and high energies. 
        Sometimes a weak QPO is also detected, making the
        distinction between this type and the previous one uncertain.
        
\item{\it type-B QPO}: a narrow transient type-B QPO (see Nespoli et al. 2003)
        is detected during
        the full observation or only during a part of it. This case will be
        discussed in detail below.

\end{itemize}

\begin{table}
\begin{center}
\caption{\small Log of the five types of power spectrum 
        in the left branch. Grouped
        observations are indicated by `--' symbols}
\begin{tabular}{l|l}
\hline
\hline
Type & Observations\\
No signal &65,68,72,73,76,77--78,83--84,89--90,100,\\
          &105--106\\
Red noise &56,57--58,60,63--64,70--71,74--75,86\\
Weak QPO  &44-46,66--67,69,79,82,85,88b,91--92,94,96,\\
          &98--99,101--102,103b,107\\
LF noise  &35,36--40,47--50,51,52--55,59,61--62,80--81,\\
          &87,88,93,95,103,104,108,109\\
type-B QPO&34,41,42,43,97\\

\hline\hline
\end{tabular}
\end{center}
\end{table}

\begin{figure}[h] \resizebox{\hsize}{!}{\includegraphics{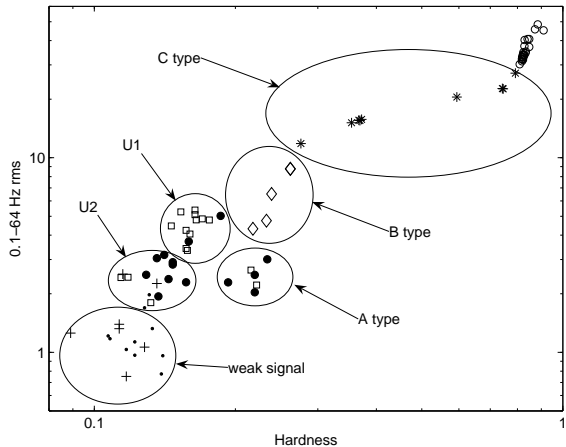}}
\caption{Plot of integrated fractional rms versus hardness for the observations
        of the right, top, and upper left branch. Different symbols 
        correspond to: right branch (open circles), top branch (asterisks),
        type-B QPO (diamonds), weak QPO (filled circles), LF noise (squares),
        red noise (plus signs) and no signal (dots). For the the definitions
        of the different elliptical regions, see Sect. 6.
} \label{rmscolore}
\end{figure}

A pattern in this complex behavior can be found by integrating the 
fractional rms in the 0.1-64 Hz range 
for all power spectra and plotting the result
as a function of hardness. The result, which includes also the points
from the right and top branches, can be seen in Fig. \ref{rmscolore}.

Here, right branch, top branch (type-C QPO) and type-B 
power spectra are clearly separated. 
Type-C points follow a correlation that extends
all the way to the right branch points, while type-B points  also follow
a correlation that deviates from the type-C one, but possibly connects to it.
A few 
observations occupy 
a specific region of the diagram, roughly on the extension of the type-B correlation.
In analogy with what is seen in XTE J1859+226 (Casella et al. 2004) we call
this region 'type A' (see below).
The other points are also grouped: we identify a group including almost
all observations with no signal, and two additional groups that we call
'U1' and 'U2'. These three groups correspond to different levels of
integrated rms (see Fig. \ref{rmscolore}). 
U1 and U2 are ketp as separate sub-groups as the power spectra of most of the
observations in U1 are classified as "LF noise", while those in U2 show
a weak QPO. This is reflected in the average power spectra (see below).
\new2{Notice that the
observations corresponding to these groups are marked with different
symbols in Fig. \ref{figure2}).}
In the following, we will 
examine the average power spectrum of these
groups, and in more detail the instances of type-B QPO detections.

\subsection{Type-B QPO}

The first detection of a type-B QPO in this outburst of GX 339--4 has already
been reported by Nespoli et al. (2003). The QPO in that observation
(Obs \#34) is transient: during the first part of the observation only a
broad bump is present in the power spectrum, 
while the narrow QPO appears suddenly
and shows marked variations in its centroid on a characteristic time scale
of $\sim$10 s (Nespoli et al. 2003). A similar QPO was present also in 
XTE J1859+226 (Casella et al. 2004).
We investigated the time evolution of the QPO of the 
other four type-B observations
by producing spectrograms with a time resolution of 4 seconds (see Nespoli
et al. 2003) and found that indeed in all four cases the QPO
is not present for the whole observation. 
The full description of the results for these observations can be found in 
Appendix A.

\begin{figure}[ht] \resizebox{\hsize}{!}{\includegraphics{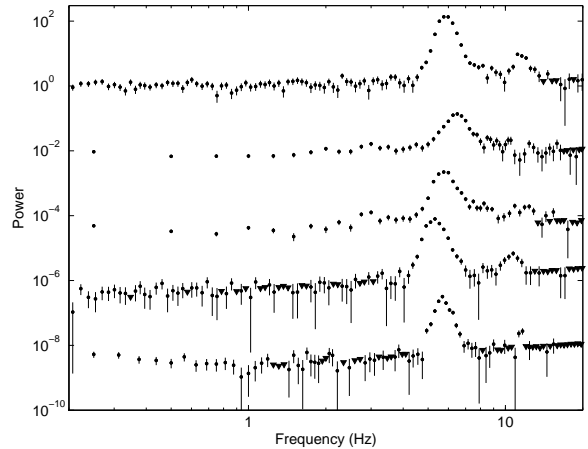}}
\caption{The Poissonian-subtracted power spectrum 
        of the five observations with a 
        type-B QPO (see text), vertically shifted for clarity. From top to bottom:
        Obs. \#34,41,42,43,97.
} \label{typeb}
\end{figure}

In summary, (transient) type-B QPOs are found in five observations
(see Fig. \ref{typeb}): their
centroid frequency ranges from 4.5 to 7.0 Hz and their fractional rms is in
the range 5--8\%. As in the case of the type-B QPO reported by Nespoli 
et al. (2003), the hardness increases slightly (by $\sim$0.02) when the
QPO is present compared to the no-QPO intervals.
Selecting QPO- and no-QPO intervals (see Appendix A), we can also notice that 
for observations \#34,43,97, the no-QPO 0.1-64 Hz rms is around 3\%, while
when the QPO is present it is around 8-9\%. For the remaining two observations,
\#41,42, the no-QPO and QPO fractional rms is always around 9\%, but as these
are the observations where the QPO appears intermittently, our selection of
the no-QPO interval was approximate and it cannot be excluded that some
residual spurious power from the QPO intervals and/or from the rise and
decay of the flux were included in the power spectra.

An example of the phase-lag spectrum for an observation showing a type-B
QPO (Obs. \#34) can be seen in Fig. \ref{lagspectra}. The errors
are large, but around the QPO frequency clear hard lags are observed,
in line with what was observed 
in previously reported cases of type-B QPOs (see e.g. Casella et al. 2004).

\subsection{Type-A QPO}

\begin{figure}[h] \resizebox{\hsize}{!}{\includegraphics{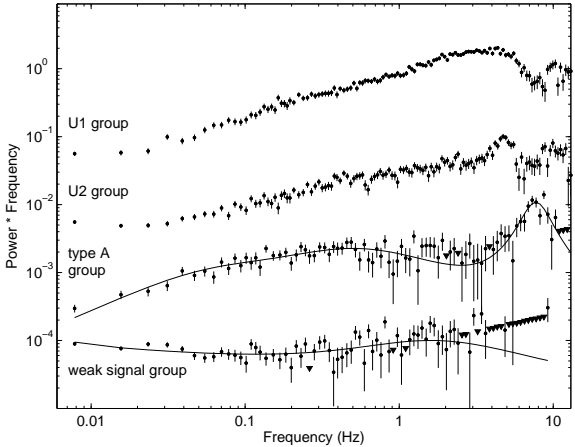}}
\caption{Average power spectra of the observations classified as (from top
        to bottom) type A, weak signal, U2 and U1 (see Fig.
        \ref{rmscolore}). The top power spectrum 
        is in power units, all the others
        are shifted downwards by a factor of 10 from each other.
        The lines are the best fit models described in the text.
} \label{typea}
\end{figure}

The average power spectrum of the observations in the 'type A' group 
(Obs. \#35,94,95,96,98--99,103b) can be seen in
Fig. \ref{typea}. It can be fitted with a model consisting of two flat-top
Lorentzians with characteristic frequencies 0.07$\pm$0.02 Hz and 
0.54$\pm$0.13 Hz respectively, plus a broad QPO with centroid 
7.49$\pm$0.18 Hz and FWHM 3.64$\pm$0.56 Hz. The integrated fractional
rms of the three components is 2.8\%, 4.7\% and 3.7\% respectively.
In the high-energy band, the power spectrum is qualitatively similar.
The broad shape could be
related to the averaging of different observations, but this shape
is remarkably similar to that observed by Nespoli et al. (2003) in the
interval before the appearance of the type-B QPO. The similarity in 
shape, centroid frequency and total rms make us identify this QPO as
of type A (see Wijnands et al. 1999; Casella et al. 2004 and references therein).

\subsection{Weak signal power spectrum}

The average power spectrum of 
the observations in the 'weak signal' group is shown
in Fig. \ref{typea}. Its total 0.1-64 Hz fractional rms is around 1.3\% and
can be fitted with a power law (index 1.16$\pm$0.03) plus a zero-centered
Lorentzian with characteristic frequency 1.75$\pm$0.33 Hz. In the high-energy
band, the power spectrum is noisy but compatible with the same shape.

\subsection{U1/U2 power spectrum}

The average power spectrum for the U1 and U2 groups appear complex (see Fig. \ref{typea}).
Since the exact shape could be the result of our averaging of different 
power spectra,
we do not present detailed fits here. The U1 power spectrum 
shows a clear band-limited
noise component,with an excess around 10 Hz. In the high-energy 
power spectrum, this
excess is clearly visible as a broad QPO. The total 0.1-64 Hz fractional
rms is 4.3\%. The U2 power spectrum is 
weaker (2.3\% rms) and a clear QPO 
is present around 5 Hz, in addition to a band-limited noise component. A
weak excess around 10 Hz is visible.
In the high-energy band, the shape of the continuum components is similar,
but the 5 Hz peak disappears, while the 10 Hz excess shows as a clear
QPO peak.

\section{The left branch (bottom section)}

 From observation \#110 (MJD 52560.41), the PCU2 count rate goes
below 560 cts/s and only weak power-law noise or no noise at all is 
detected in the power spectrum. From Fig. \ref{figure2}, it is evident that all
these observations are at relatively low hardness, and a much smoother
path is followed mostly monotonically downwards, in the HID.
We therefore accumulated one average 
power spectrum from these observations 
(\#110 to \#148f). The power spectrum is well described by
a single power law with slope 0.82$\pm$0.02. The total integrated rms 
between 0.1 and 64 Hz is only 1.6\%. No high-frequency cutoff is observed,
with a lower limit of 25 Hz.
The average power spectrum along this branch is 
similar to that of the weak-signal
power spectrum described above, although the power-law here is slightly flatter.

\section{The bottom branch: back to the start line}

The last observation on the left branch, with no significant signal in 
he power spectrum is from MJD 52693.73 (Obs \#148f). Just before this date 
there is a considerable hardening (see Fig. \ref{figure1} and
\ref{figure2}). In the following observation (\#149, MJD 52694.92),
band-limited noise, often with a significant QPO,
appears once again in the power spectrum. The transition is marked with a
dotted line in Figures \ref{figure1} and \ref{figure2} and corresponds
roughly to a hardness of 0.2.

\begin{figure}[h] \resizebox{\hsize}{!}{\includegraphics{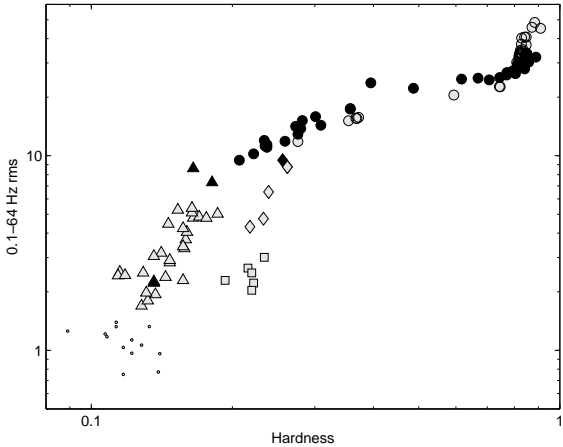}}
\caption{Plot of integrated fractional rms versus hardness for the observations
        of the bottom. Open symbols represent 
        the same points as in Fig. \ref{rmscolore}: right branch and type C
        (circles), diamonds (type B), triangles (U1/U2), squares
        (type a), and dots (weak signal). 
        Filled symbols correspond to different power-spectrum 
        shapes observed in the bottom branch (see text): flat-top noise
        (circles), weak noise (triangles), 1Hz QPO (diamond). 
} \label{rmscolore2}
\end{figure}

The hardening is almost monotonic throughout the bottom branch, 
with the exception of a few observations 
that indicate a short period of softening. During three of these 
softer observations
(Obs. \#154), the hardness went back below 0.2 and no noise components were
detected in the power spectrum, 
indicating that the source was back in the left-branch
region. For all
other observations after the dotted line, noise was observed 
in the power spectrum.
The evolution of the power spectrum can 
be followed once more by examining a plot
of integrated fractional rms versus hardness as in Fig. \ref{rmscolore}.
Figure \ref{rmscolore2} shows this diagram, where black symbols indicate 
observations from the bottom branch and gray symbols show the same points
as in Fig. \ref{rmscolore}.

All observations indicated with circles show flat-top noise in the 
power spectrum, 
sometimes with a QPO. The three observations with hardness below 0.2
show very weak noise. One observation, located in the `type B' region from 
the left branch, shows a very peculiar power spectrum 
consisting essentially of a
single peak at 1 Hz. These three classes are discussed

\subsection{Weak noise power spectra}

These three observations with hardness below 0.2 (Obs. \#153c,154,154b)
show very little power so that it is difficult to determine the shape
of the power spectrum at these low count rates. The upper limits on the rms are
consistent with the value from Sect. 7.

\subsection{1Hz QPO}

\begin{figure}[h] \resizebox{\hsize}{!}{\includegraphics{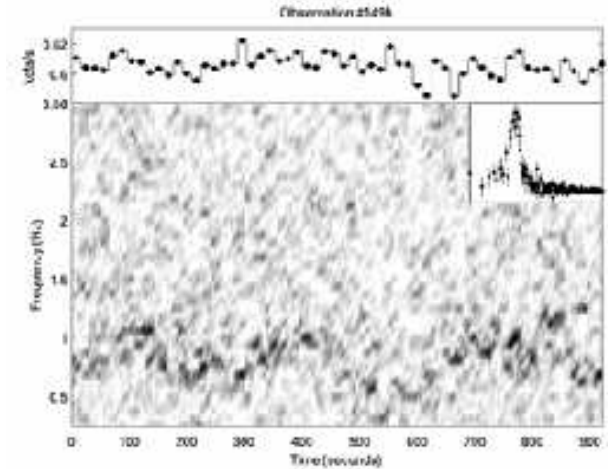}}
\caption{Top panel: 900s of the light curve of Obs. \#149h (16 second
bin size). Bottom panel: corresponding spectrogram (time integration 16 seconds,
time step 4 seconds), where darker points correspond to higher power. Inset:
average power spectrum from the full observation.
} \label{pds1Hz}
\end{figure}

Observation \#149h, on MJD 52707.91 appears completely different from the
surrounding ones. The power spectrum and the corresponding spectrogram are 
shown in Fig. \ref{pds1Hz}. The power is dominated by a single QPO peak 
at 0.91$\pm$0.04 Hz, with a FWHM of 0.33$\pm$0.06. The spectrogram 
shows that this peak is not stable in time and shifts from 0.7 to 1.2 Hz
over the observation. Aside from the frequency, this behavior is similar
to that of the type-B QPOs described above, including the position in the
color-hardness diagram. We extracted the QPO phase
lag as described in the previous sections, but the results is
consistent with no lags (-0.077$\pm$0.076 rad). The fractional rms in the
QPO peak is $\sim$7\%.

\subsection{Flat-top noise power spectra}

\begin{figure}[h] \resizebox{\hsize}{!}{\includegraphics{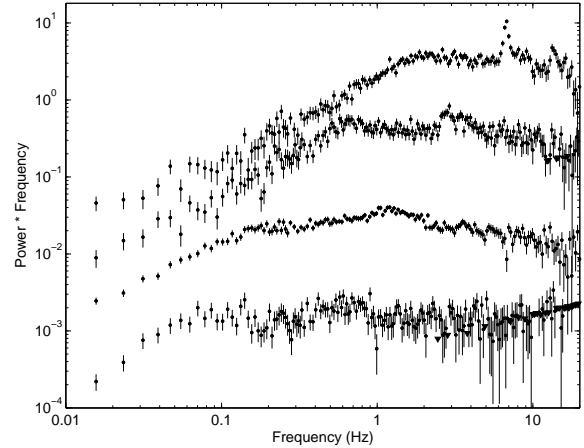}}
\caption{Four representative power spectra 
        from the bottom-branch observations. The time sequence
        here is bottom to top: observations \#161,158b,155,149g.
        The power spectra are plotted in the $\nu P_\nu$ representation 
        (Belloni et al. 1997).
} \label{bottompds}
\end{figure}

All other observations of the bottom branch, which at the end bends downwards
and becomes vertical, joining the initial right branch, show rather strong
noise components. Some examples are shown in Fig. \ref{bottompds}.

\begin{table}
\begin{center}
\caption{\small Main characteristic frequencies (in Hz) for the bottom branch of the HID}
\begin{tabular}{lccccc}
\hline
\hline
Obs \# & $\nu_1$     &  $\nu_2$         &   $\nu_3$     & $\nu_{Q}$  \\
\hline
\multicolumn{5}{c}{Bottom branch}\\
\hline

149 &2.91$\pm$0.21    & ...             &        ...       & 8.59$\pm$0.18    \\
149b&1.87$\pm$0.10    & 5.42$\pm$0.80   &        ...       & 8.02$\pm$0.04    \\
149c&2.58$\pm$0.19    & ...             &        ...       &     ...          \\
149d&3.24$\pm$0.44    & ...             &        ...       &     ...          \\
149e&3.13$\pm$0.28    & ...             &        ...       &     ...          \\
149f&3.19$\pm$0.49    & ...             &        ...       & 6.72$\pm$0.06    \\
149g&3.39$\pm$0.22    & ...             &        ...       & 6.77$\pm$0.02    \\
149h&    ...          & ...             &        ...       &     ...          \\
151 &2.71$\pm$0.11    & ...             &        ...       & 8.03$\pm$0.06    \\
150 &3.04$\pm$0.24    & ...             &        ...       & 8.10$\pm$0.13    \\
152 &4.00$\pm$0.57    & ...             &        ...       & 7.77$\pm$0.09    \\
152b&2.82$\pm$0.37    & ...             &        ...       &     ...          \\
153 &3.33$\pm$0.21    & ...             &        ...       &     ...          \\
153b&3.84$\pm$0.56    & ...             &        ...       &     ...          \\
153c&    ...          & ...             &        ...       &     ...          \\
154 &    ...          & ...             &        ...       &     ...          \\
154b&    ...          & ...             &        ...       &     ...          \\
154c&3.37$\pm$0.29    & ...             &        ...       &     ...          \\
154g&3.01$\pm$0.26    & ...             &        ...       &     ...          \\
154d&3.13$\pm$0.53    & ...             &        ...       &     ...          \\
154e&3.81$\pm$0.49    & ...             &        ...       & 6.12$\pm$0.10    \\
154f&1.09$\pm$0.09    & 6.64$\pm$1.20   &        ...       & 4.74$\pm$0.07    \\
155 &0.67$\pm$0.06    & 4.22$\pm$0.86   &        ...       & 2.85$\pm$0.07    \\
155b&0.44$\pm$0.08    & 2.86$\pm$0.48   &        ...       &     ...          \\
155c&0.48$\pm$0.21    & 2.53$\pm$0.87   &        ...       &     ...          \\
156 &0.22$\pm$0.02    & 1.90$\pm$0.10   &        ...       &     ...          \\
156b&0.24$\pm$0.03    & 1.45$\pm$0.20   &        ...       &     ...          \\
157 &0.18$\pm$0.03    & 1.48$\pm$0.08   &        ...       &     ...          \\
157b&0.27$\pm$0.03    & 1.90$\pm$0.44   &        ...       &     ...          \\
158 &0.20$\pm$0.05    & 1.33$\pm$0.19   &        ...       &     ...          \\
159 &0.22$\pm$0.02    & 1.15$\pm$0.08   &        ...       &     ...          \\
158b&0.28$\pm$0.02    & 1.22$\pm$0.05   & 6.13$\pm$0.56    &     ...          \\
159b&0.23$\pm$0.02    & 1.17$\pm$0.06   & 6.31$\pm$0.69    &     ...          \\
159c&0.14$\pm$0.05    & 1.25$\pm$0.02   &        ...       &     ...          \\
160 &0.36$\pm$0.06    & 2.44$\pm$0.24   &        ...       &     ...          \\
161 &0.14$\pm$0.03    & 1.40$\pm$0.39   &        ...       &     ...          \\
161b&0.13$\pm$0.04    & 0.85$\pm$0.21   &        ...       &     ...          \\
162 &0.12$\pm$0.03    & 2.11$\pm$0.58   &        ...       &     ...          \\
162b&0.23$\pm$0.05    & 2.92$\pm$1.29   &        ...       &     ...          \\
163 &0.11$\pm$0.01    & 1.84$\pm$0.32   &        ...       &     ...          \\
162c&0.18$\pm$0.03    & 3.17$\pm$0.18   &        ...       &     ...          \\
163b&0.13$\pm$0.03    & 1.21$\pm$0.04   &        ...       &     ...          \\
163c&0.23$\pm$0.06    & 1.85$\pm$0.62   &        ...       &     ...          \\
164 &0.06$\pm$0.01    & 1.33$\pm$0.31   &        ...       &     ...          \\

\hline
\end{tabular}
\end{center}
\end{table}

\begin{figure}[h] \resizebox{\hsize}{!}{\includegraphics{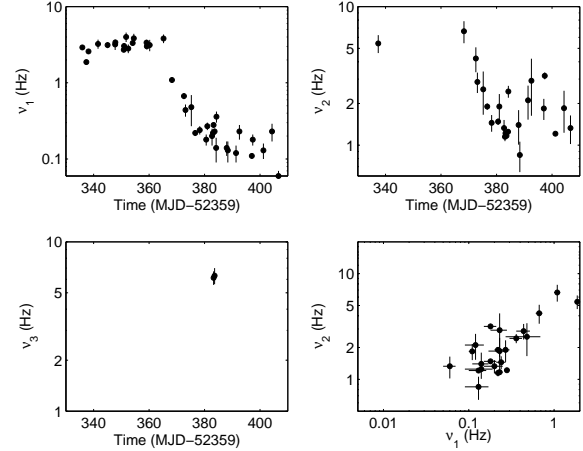}}
\caption{Same as Fig. \ref{frequencies}, but for the bottom-branch
        observations.
} \label{figure4bottom}
\end{figure}

We fitted the power spectra with the same model adopted for the 
top-branch and right-branch observations
at the beginning of the outburst. Only a maximum of three components were
needed (see Table 4
\new2{and Fig.\ref{figure4bottom}}). 
A narrow QPO was found to be significantly present in 10
observations. The first observations, with one exception, show only one flat-top
noise component. When the characteristic frequencies become low, a second noise
component appears. Only in two cases it was possible to detect a third broad
component.
Unlike the case of the detections summarized in Table 2, the QPO frequency
is substantially higher than the characteristic frequency of the first noise
component, suggesting that \new2{the latter}
should be identified with $L_b$ and not with 
$L_h$ as in the previous case. This is confirmed, although with some deviations,
by the plot in Fig. \ref{topbottom}, which shows the corresponding WK correlation.
\new2{The three noise components $L_1$,$L_2$ and $L_3$ in Table 4 
are therefore identified as
$L_b$, $L_h$ and $L_\ell$.} The latter of these three 
identifications can only be tentative, as it
is not possible to check other correlations for lack of appropriate pairs
of frequencies.
\new2{Comparing Fig. \ref{frequencies} and Fig. \ref{figure4bottom} one can
see that an alternative identification could be the same as for the top
branch.
The evolution of $\nu_1$ as a function of count rate and hardness for the
bottom branch is shown in Fig. \ref{nu_rate_color} (empty circles).
Clearly, it is not possible to obtain a firm identification for these
components.}

The integrated fractional rms also increases with increasing characteristic
frequencies: the L$_b$ component goes from 8\% up to 31\% at the last
observation, while the L$_h$ component goes up to 23\%, values similar to
those at the beginning of the outburst. The L$_{Q}$ component has an integrated
fractional rms of 4-8\%.
For these observations at low count rate, the signal is too weak to measure 
phase-lags significantly different from zero.

\section{High-frequency QPOs}

No significant high-frequency peaks were seen in the single power spectra, 
either at
low or at high energies. Since high-frequency QPOs in other systems have
been detected when the source was in the IMS/VHS state, we accumulated average
high-energy power spectra from all 
type-B and type-A observations and searched for
excesses above 30 Hz. Fits were performed to the
power spectrum with a model 
consisting of a flat level (to account for the Poissonian
component) plus a Lorentzian. The search 
was done fixing four values of $Q$: 5, 10, 15 and 20, and with different 
starting centroid frequencies in order to make sure the best fit was
reached. No significant peak was detected. The $3\sigma$ one-trial upper limits
for type A are 2.6\%, 2.3\%, 2.0\% ad 1.3\% for the four values of $Q$.
For type B they are 2.6\%, 2.3\%, 2.0\%, 2.8\%.


\section{Discussion}

\begin{figure}[h] \resizebox{\hsize}{!}{\includegraphics{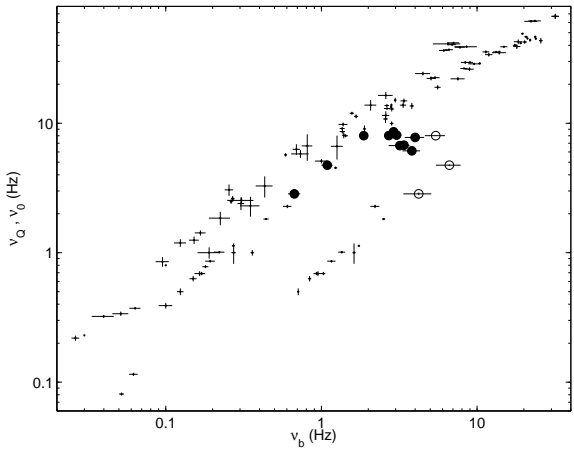}}
\caption{WK correlation (see Wijnands \& van der Klis 1999) for the observations
        on the bottom branch of the HID (filled circles for $\nu_b$ and empty
        circles for $\nu_h$). The small points are from 
        Belloni, Psaltis \& van der Klis (2002). 
} \label{topbottom}   
\end{figure}

\subsection{The outburst}

The results presented above outline a rather clear picture for the 
outburst of GX 339--4. From the combined timing and color properties,
we can identify four main states, which have a correspondence with
those determined from a detailed spectral analysis (Homan et al.
in prep.) and optical/nIR data (Homan et al. 2005a).  The transitions
between these states correspond to precise locations in the HID and
can be very fast. In the following we examine these states in terms
of their timing features.

\subsubsection{The right branch: Low/Hard State (LS).}
The observations  on
the right vertical branch  in Fig. \ref{figure2} show hardness and
timing properties typical of the  ``canonical'' LS.  The PSD is
dominated by strong band-limited noise (see top four power spectra in  Fig.
\ref{figure3}). Notice the similarity between the second power spectrum from the
top in Fig. \ref{figure3} and that shown by Belloni et al. (1999) for the
LS of GX 339-4.
This noise can be decomposed into a relatively small
number of Lorentzian components, which have clear counterparts in the
power spectra of other systems (see Nowak 2000; Belloni, Psaltis \& van der
Klis 2002). As usual in black-hole transients, this corresponds to
the earlier part of the outburst, where the X-ray flux increases
steadily (see Cui et al. 1999).  Interestingly, the right branch in
the HID is almost vertical, indicating that the spectral shape does
not change much during the rise (see Vignarca et al. 2002; Homan et
al., in prep.). In a few of the power spectra along this branch, a low-frequency
QPO is observed (see Grebenev et al. 1999; Belloni et al. 1999).
All characteristic frequencies of the noise
components increase as the source brightens. Hard lags increasing
with time are observed along this branch, both for the  continuum and
the QPO (Figs. \ref{lags} and \ref{lagspectra}, see also Nowak, Wilms
\& Dove 1999).

The origin of the strong aperiodic variability observed in this state
is not known. Our results confirm that in transient systems the
Low/Hard state is associated to the early and late stages of the outburst,
\new2{but not necessarily with low luminosities}.
The observed
time evolution of flux and characteristic frequencies, both at the
beginning and at the end of the outburst, suggests that they are correlated
with mass accretion rate. The similarity in timing and color parameters
indicate that the outburst ends with similar physical conditions 
similar to the
beginning more than one year earlier.

\subsubsection{The top branch: Hard Intermediate State (HIMS).}
 From observation
\#25 (MJD 52398), the hardness starts to decrease at a faster rate
and the source enters the top branch in the HID (see Fig.
\ref{figure2}). This horizontal path in  the HID is followed
considerably faster in time than the LS branch. This softening is associated
with an increase of the flux from a soft thermal component, with a
simultaneous steepening of the hard component (Homan et al. in prep.).
As mentioned above, the transition from the right to
the  top branch corresponds to a marked change
in the behavior of the timing parameters (see Fig. \ref{frequencies})
and at the same time it corresponds to a clear change in the
properties of the IR/optical/X-ray correlations (Homan et al.
2005a). 

During this period,  a clear type-C QPO appears in the power 
spectrum, which can be linked
to that detected in some observations of the right branch.
The overall noise,  however, can still be fit with the same
components, whose characteristic frequencies increase faster with
time than on the right branch, and whose fractional rms is reduced (see Fig.
\ref{frequencies}). The continuum and QPO phase lags are still
positive (hard lags soft) and higher than in the LS (see Figs.
\ref{lags} and \ref{lagspectra}). This provides a clear
link between this branch and the LS. This power-spectrum shape has been
traditionally  observed in the Very High State and Intermediate State
of BHCs (see Belloni et al. 1997; M\'endez \& van der Klis 1997),
starting from the early observations of GX 339--4 and GS 1124--683 with
Ginga (Miyamoto et al. 1991,1993,1994), when these timing properties
could be associated to the presence of a  relatively strong
contribution from the hard component, while at softer X-ray colors, a
different power spectrum was observed 
(see below). The power spectrum of the C state of
GRS 1915+105 are clearly associated to this state (see Morgan et al.
1997; Trudolyubov 2001: Reig et al. 2000). Notice that along this
branch, the source does not reach the soft  vertical branch yet (see
Fig. \ref{figure2}). The end of this branch is again marked by a
clear transition in the  timing parameters. 

We call this branch the Hard Intermediate State (see also
Homan \& Belloni 2005). The fact that the timing components
appear to be the same as some of those observed in the Low/Hard State
suggests a similar physical origin, while the higher values for the
characteristic frequencies are probably associated to a smaller value
of the radius of the accretion flow associated to this time scale. 
This indicates that the same spectral
component is responsible for this variability (see also Homan et al.
in prep.).

\subsubsection{The Top/left branch: Soft Intermediate State (SIMS).}
 From Obs.
\#34, as the source moves further left on the top branch, the timing
properties of  GX 339--4 change abruptly. On Obs. \#34, no flat-top
noise component is detected, and a type-B QPO appears at $\sim$6 Hz
(Nespoli et al. 2003). It is clear that a rather sharp transition
took place here. Notice that  the difference in hardness between Obs.
\#33 and \#34 is small.  From this observation on,  GX 339--4 moves
irregularly in the HID, remaining in the color range 0.05-0.25 (see
Figure \ref{figure2}). No type-C QPO is detected along this branch,
although we cannot at present exclude that the U1/U2 QPOs are of
type C.

This branch is not as straight and monotonic as the previous ones, as the source moves up
and down in flux by a large amount (see Fig. \ref{figure1}), and also 
its hardness does span a rather extended range in log space.
Here, after grouping of observations together (see Sect. 6), we can identify
a clear dependence of timing properties on hardness (see Fig. \ref{rmscolore})
Above HR=0.27, the power spectra show type-C QPOs, 
in the HR range 0.19-0.27 type-A and
type-B QPOs, below HR=0.19 are the U1/U2/no-signal power spectra. 
All these features
are observed over a large range of count rates, indicating that the spectral
shape and not source flux is what determines the timing properties. 
Figure \ref{rmscolore} indicates that the type A/B/C
oscillations also correspond to different non-overlapping ranges of 
total 0.1-64 Hz rms.
Notice that in the five observations where type-B QPOs are observed, the
QPO is not present all the time. In the intervals without type-B QPO,
in the observations where the extraction of no-QPO power spectra 
was not problematic,
the no-QPO hardness is slightly lower and the corresponding integrated
fractional rms drops to $\sim$3\%, a value that brings these observations
in the type-A region of Fig. \ref{rmscolore}. The identification of these
intervals with type A is strengthened by the detection of a $\sim$8 Hz
QPO in some of those power spectra.

The characteristics of the type-B QPOs
have been observed before by Ginga in the power spectrum of
GS 1124--683 and GX 339--4 (see Miyamoto et al. 1991; Takizawa et al.
1997). Also fast transitions between two different types of power spectrum, 
corresponding
to type-C and type-B QPOs, have been observed in those systems (being part
of the original definition of Very Hard State), together
with spectral changes. Fast transitions have been observed
with RXTE in the bright transient XTE J1859+226 (Casella et al. 2004). Both
type-B/type-A and type-B/type-C transitions were observed: here a very 
sharp threshold in count rate was observed to correspond to the transitions.
Recently, the transition between the top and the top/left branch has been 
identified as
marking the time of the launch of relativistic outflows (see Fender, Belloni
\& Gallo 2004). All these properties, together with the properties of the
energy spectra of GX 339--4, lead to the conclusion that the two branches
correspond to two different physical states of accretion.
We call this top/left branch Soft Intermediate State (see Homan \& Belloni
2005). Notice that this state, unlike the others,
is characterized by a collection of different properties in the power spectra

Finally, softer observations along this branch are more difficult to 
classify. Our U1/U2/weak subdivision is the result of extensive
averaging, due to the shortness of most observations. 

\subsubsection{The Bottom/Left Branch: High/Soft state}

Below a PCU2 count rate of 500 cts/s, corresponding to MJD$>$52559, only 
weak variability is observed. Looking at Fig. \ref{figure2}, one can see
that from this time on the movement in the HID is monotonically decreasing
in count rate, with a general softening. 
We identify this branch with the canonical High/Soft State. A long RXTE/PCA
observation to GX 339--4 in this state during a previous transition showed
a similar power spectrum with a slightly 
flatter power-law slope 0.62 (Belloni et al.  1999).
Notice, however, that the observations of the upper left branch 
labeled ``weak signal'' in Fig. 
\ref{rmscolore} (see Sect. 6.3) have properties that are also consistent
with the High/Soft State

\subsubsection{The bottom branch: back to the Hard-Intermediate and Hard states}

The bottom branch corresponds to a final hardening of the spectrum. As
one can see from Fig. \ref{figure2}, the hardening starts on the last 
observation of the left branch, but clear timing changes are observed
only when the hardness goes above $\sim$0.2.
What we called bottom branch is clearly complex. The first part of the
branch is horizontal: the source moves to the right, although it makes
a significant excursion back to the region HR$<$0.2. In this region, once
more the power spectrum shows little noise level in the 
form of a power law, indicating
that it was a brief interval back into the High/Soft state. In the rest of
the branch however, the power spectrum shows 
clear flat-top band-limited noise whose
characteristic frequencies follow the same correlations as those from the
top and right branch. Also, Fig. \ref{rmscolore2} indicates that these
power spectra show indeed the properties of the 
Hard Intermediate and Low/Hard States and fill the gaps in Fig.\ref{rmscolore},
suggesting a possible 
connection with the U1/U2 points (see Fig. \ref{rmscolore2}).
The major difference is that here the $L_b$ component clearly appears in 
the power spectrum, so that the source follows the main WK correlation (see
Fig. \ref{topbottom}).

Notice that the bottom branch bends smoothly and becomes vertical, 
ending up parallel to the initial right branch. However, here we do
not find a sharp transition in some parameter and do not know where
to put the transition to the Low/Hard state, which clearly takes
place. It is interesting however to note that around MJD 52740 
\new2{Obs. \#157b)} two things
take place at the same time: the source reaches the same hardness value
at which it left the LS (see Fig. \ref{figure1}), 
and the J-band light curve clearly shows a change (see Bailyn \& Ferrara 2004).
\new2{We identify this date as the marking of the transition to the Low/Hard State
(see dotted line in Fig. \ref{figure2}).}
Finally, our last  point is almost exactly in the same
location on the HID as our first point.

It it important to notice that, unlike what happened at high count
rate in the early part of the outburst, in moving from the High/Soft
state to the Hard Intermediate State, there are no observations showing 
clear timing features that would indicate the Soft Intermediate state, such as
for instance type-A or type-B QPOs.
Interestingly, there is one
observation (Obs. \# 149h)  which shows completely different timing
properties: here the power is dominated by one single QPO peak at
$\sim$1 Hz, which shows rather large variations in time (see Fig.
\ref{pds1Hz}). Although no significant phase-lag  estimate could be
obtained, we could speculate that this is a variant of the SIMS
observed at low luminosities. Clearly, more
observations of this type would be needed in order to confirm this. A
detailed analysis of this transition in a number of BHC is presented
by Kalemci et al. (2004), who also conclude that the timing
properties change more abruptly than spectral (and therefore
hardness) properties. A 1-Hz QPO with similar properties was observed from 
XTE J1650--500 (Rossi et al., in preparation), and is also present
in RXTE archival observations of GX 339--4 

\subsection{Source states}

In summary, the picture that emerges from the 2002/2003 outburst of 
GX 339--4 (see Fig. \ref{figure2}, Homan et al. in prep. and Homan 
\& Belloni 2005) is
consistent with  that coming from a full spectral analysis and from
multiwavelength correlations (Homan et al. 2005a). Four main regions
are identified in the HID, corresponding to four separate states. 
These states and their transitions can be identified from the changes
in the parameters of the fast time variability. The Low/Hard State
(the right branch) corresponds to the first phase of the outburst and
shows little spectral variations over a rather wide range of PCA
count rate. Its timing properties are consistent with those usually
seen in this state in GX 339--4 and in other systems. 

The High/Soft State (the left branch) has a much lower hardness and
small amplitude fast variability. More interesting is what happens in
between these two  branches, which in our data 
clearly corresponds to transitions between the two states mentioned above.
The evolution of timing properties in the  Hard Intermediate State (the
top and bottom branches) can be seen as a continuation of that in the
Low/Hard State and is rather similar for the high-flux and low-flux
branches, corresponding to opposite transitions. A clear marker for
the hard-to-soft transition can be found in the timing parameters as
well as in the IR/X-ray correlations, but the situation seems to be
smoother for the soft-to-hard transition. Notice that this
transition appears to be reversible, as for some time GX 339--4 goes
back to the High/Soft state before continuing to harden to reach the
Low/Hard State. During this state, type-C QPOs are observed. The
phase lags of the type-C QPOs observed here are always positive, even
when the centroid frequency of the QPO approaches 10 Hz. This is in
contrast to what observed in  GRS 1915+105 and other transient
systems (see Casella et al. 2004), indicating that the phase-lag
behavior cannot be categorized in a simple way (notice that the
contribution of the continuum to the phase lags can complicate its
measurement, and along the top branch the $L_h$ component dominates
it, unlike in other systems).

A fourth region (or branch) in the HID can be identified (the
top/left branch), possibly present only in the high-flux part of the
HID. Here the timing properties are markedly different from those in
the Hard Intermediate State,  indicating a transition to one of the collective
properties which we call Soft Intermediate State. 
Type-A and type-B QPOs appear and show rapid variations and
transitions on short time scales. The transition to this state is
very sharp and, unlike in XTE J1859+226, no more type-C QPOs are
observed after it (see Casella et al. 2004) until the final part of
the outburst. Interestingly, the
transition into this state appears to be associated with the
ejection of relativistic jets (Gallo et al. 2004;  Fender, Belloni \&
Gallo 2004), which again makes it an important state to study.  Up to
now, it has been observed in a number of systems: GX 339--4 (Miyamoto
et al. 1991,1993; Takizawa et al. 1997; Nespoli et al. 2003; this
work), GS 1124--683 (Miyamoto  et al. 1994), XTE J1550--564 (Homan et
al. 2001; Remillard et al. 2002),  XTE J1859+226 (Casella et al.
2004), H1743--322 (Homan et al. 2005b), and 4U 1543--47 (Park et al.
2004), making it a rule rather than an exception. Our work shows that
the behavior in this state is more complex, and additional
unclassified QPOs appear (U1/U2, see above). Power spectra with similar
features have been observed in other systems such as XTE J1550--564
(Remillard et al. 1999), 4U 1630--47 (Dieters et al. 2000;
Trudolyubov, Borozdin \& Priedhorsky 2001) and possibly GRS 1915+105
(Morgan et al. 1997; Trudolyubov 2001). Clearly, these features
deserve being studied in more detail with longer observations.

Interestingly, a type-B QPO has not been
observed up to now in GRS 1915+105, despite the wide range of
characteristics shown by this system (see Fender \& Belloni 2004
for a review). Fender, Belloni \& Gallo (2004)
interpret the X-ray/radio correlations of GRS 1915+105 and other transients
in the framework of the same model; however, the time scales involved in 
GRS 1915+105 are much shorter and it is possible that the source moves
too fast through its HID to show the timing features of this state.
A full comparative analysis of a number of transient systems will be 
presented in a forthcoming paper.

The state paradigm presented here is not the same as that outlined in 
McClintock \& Remillard (2005). Although the Low/Hard and High/Soft state
 can obviously 
be identified with the Hard and Thermal-Dominant states, our two 
additional states
are not the same as the Steep-Power-Law and 
Intermediate-State of McClintock \& Remillard (2005). A precise comparison
is not possible, since their classification is based on model-dependent
parameters coming from spectral fits, but the variety of power spectra shown to 
be associated to the SPL in McClintock \& Remillard (2005), which includes
both type-A/B and type-C QPOs, excludes a one-to-one relation to ours.

\subsection{Quasi-Periodic Oscillations}

With these results, the A/B/C classification of low-frequency QPOs in 
black-hole candidates is strengthened. There are still undecided cases
such as our U1/U2 power density spectra, but they all correspond to 
low signal-to-noise signals that might turn out to belong to one of the
three types once observed with higher statistics.

Type-C QPOs are the most common and are observed in all systems that 
show LS and HIMS (see McClintock \& Remillard 2005; van der Klis 2005).
Their properties are rather well defined. They are clearly associated 
to the hard component in the energy spectrum (see e.g. Casella et al. 2004) 
and appear during states when the hard component shows a high-energy cutoff
and is possibly associated to an outflow (see Fender, Belloni \& Gallo 2004).
Their frequency is strongly correlated with other observed properties such
as the source flux and X-ray color. They are associated to the QPOs 
observed sporadically in the LS, and a similar association can be made 
between the broad-band noise properties of HIMS and LS. There is no
consensus as to the physical origin of type-C QPOs, but it is clear
that these oscillations are particularly important for our understanding
of the accretion flow, as they relate to a spectral component that could
be associated to an outflow from the system (see Kanbach et al. 2001; 
Rodriguez et al. 2004).

QPOs of types A and B are detected over a much smaller range of frequencies
(see also Casella et al. 2004) and, while having a similarly hard
energy spectrum, are most likely associated to a different hard component, 
for which no high-energy cutoff is observed (see Zdziarski et al. 2001).
Rodriguez et al. (2004), using simultaneous RXTE/Integral observation,
showed that for GRS 1915+105, the energy dependence of the type-C QPO
indicates the presence of a high-energy cutoff, while the energy spectrum 
is composed of two separate components.
These type-A/B oscillations appear in soft intervals of the
HID, where no compact jet emission is observed in the radio (Fender, Belloni \&
Gallo 2004).
The fast transitions observed between these two types of QPO and in 
particular transitions involving also the type-C QPO (Casella et al. 2004)
could be indicative of short transitions to the HIMS.
These mini-transitions could be responsible for small jet
ejections similar to those observed on short (hour) time scales in 
GRS 1915+105 (Fender \& Belloni 2004; Fender, Belloni \& Gallo 2004).
These transitions are indeed related to variations in hardness (GX 339--4)
or intensity (XTE J1859+226). 
Their limited range of frequencies, especially of the type-B QPOs is
particularly interesting. Since a transient oscillation at a similar
frequency has been observed in 
the  neutron star system 4U 1820--30 (Belloni, Parolin \& Casella 2004), if
this association is correct the frequency of type-B QPOs should have a weak
dependence on the mass of the compact object. Notice that the 1 Hz QPO 
detected near the end of the outburst appears at a hardness value of 0.26,
comparable to that at which the 6 Hz type-B QPOs are seen. Since the 
properties of this 1 Hz oscillation (besides its centroid frequency)
are compatible with those of type-B QPOs, it is possible that the small
range of the frequency of these QPOs is due to their appearing only at
relatively bright phases of the outburst. In other words, there is also
a count rate dependency of the frequency of type-B QPOs, but most of these
features appear at a high count rate. In our data, there is some evidence
of a count rate dependence of the 6Hz QPO, as it is in the case of
XTE J1859+226 (Casella et al. 2004).
At any rate, the fast transitions between QPO types observed here (and in 
other systems) indicate that state transitions take place on very short
time scales and are clearly traced by the timing properties. The three
types of QPO described here (A/B/C) 
are rather different and are possibly tracers
of separate spectral components.

\section{Summary and conclusions}

We analyzed a large set of RXTE/PCA observations of GX 339--4 during its
complex 2002/2003 outburst. From the timing and color results, we identify
four main states of the source, with a distinctive pattern of evolution of
the outburst. 
The timing characteristics during the outburst are complex, but show
simultaneous spectral transitions that allow a clear categorization.
These states (in time sequence) are:

\begin{itemize}

\item {\it Low/Hard state}: \new2{this state is associated with relatively low
values of the accretion rate, i.e. lower than in the other bright states, 
although it is clearly not limited to low flux intervals. The
energy spectrum is hard and the fast time variability is dominated by a
strong ($\sim$30\% fractional rms) band-limited noise. Sometimes, low
frequency QPOs are observed. The characteristic frequencies detected in the
power spectra follow broad-range correlations (see Belloni, Psaltis \& van der
Klis 2002).
} 
        
\item {\it Hard Intermediate state}: \new2{in this state, the energy spectrum is
softer than in the LS. The
power spectra feature band-limited noise with characteristic frequency
higher than the LS and a rather strong 1-8 Hz type-C QPO. The
frequencies of the main components detected in the power spectra
extend the broad correlations mentioned for the LS. However, a sharp 
transition from the LS can be identified by looking at the IR/X-ray 
correlation (Homan et al. 2005a).
}

\item {\it Soft Intermediate state}: \new2{here the energy spectrum is systematically
softer than the HIMS. No strong
band-limited noise is observed, but transient type-A and type-B QPOs appear, the
frequency of which spans only a limited range. The transition from the HIMS
is very sharp and marked by the transition from type-C to type-A/B QPOs in 
the power spectra, despite the relatively small changes in hardness.
This state is characterized by a variety of timing properties, of which 
type-A/B QPOs are only a subset. 
}

\item {\it High/Soft state}: \new2{the energy spectrum is very soft. 
Only weak power-law noise is observed
in the power spectrum. The onset of this state is identified with the
systematic non-detection of features in the power spectrum.
}

\end{itemize}

\begin{figure}[h] \resizebox{\hsize}{!}{\includegraphics{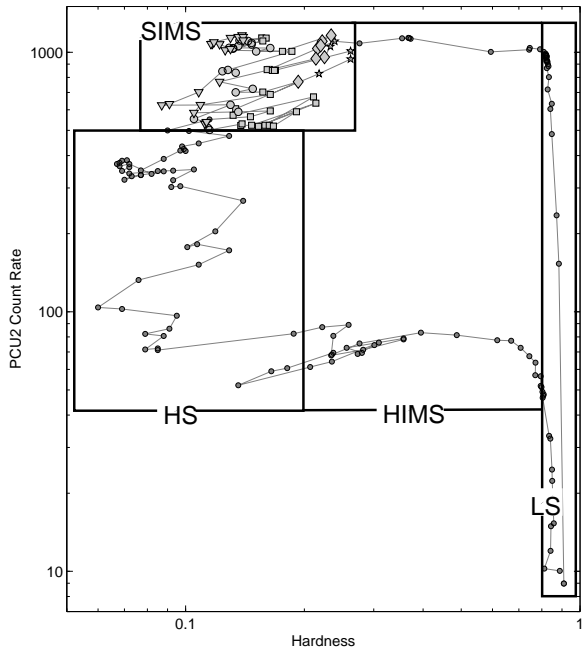}}
\caption{Same Hardness-Intensity as in Fig. \ref{figure2}, but areas
marking
the regions corresponding to the source states described in the text.
} \label{final_hid}
\end{figure}

\new2{
After the HS, a transition back to the HIMS is observed, marked by the
sudden appearance of band-limited noise with type-C QPOs. No clear evidence
of a SIMS is observed in these later stages of the outburst, although one
observation shows a strong 1-Hz QPO which could be of type B. The HIMS
evolves into a LS, roughly following an evolution opposite of that of 
the beginning of the outburst. 
These four states are observed in other systems and they are clearly separated
by sharp transitions that are essential for their identification.
Their location and boundaries on the HID of GX 339-4 are shown in 
Fig. \ref{final_hid}.
} 

The phenomenology of the states, their evolution and the 
transitions between them
could constitute a firm starting point for the development of physical
models for the outbursts of black-hole transients.

\begin{acknowledgements}
This work was partly supported by INAF-PRIN 2002 grant and MIUR-PRIN grant
2003027534\_004.
We thank the anonymous referee for constructive comments.

\end{acknowledgements}

\appendix

\section{Type-B QPOs in detail}

In this Appendix, we describe the detailed analysis of the four type-B
observations found in addition to that reported by Nespoli et al. (2003).
Notice that the fits to average
power spectra described below are only indicative, since it is clear that there
are major fast variations in the QPO parameters on rather short time scales
(see also Nespoli et al. 2003).

\begin{figure}[h] \resizebox{\hsize}{!}{\includegraphics{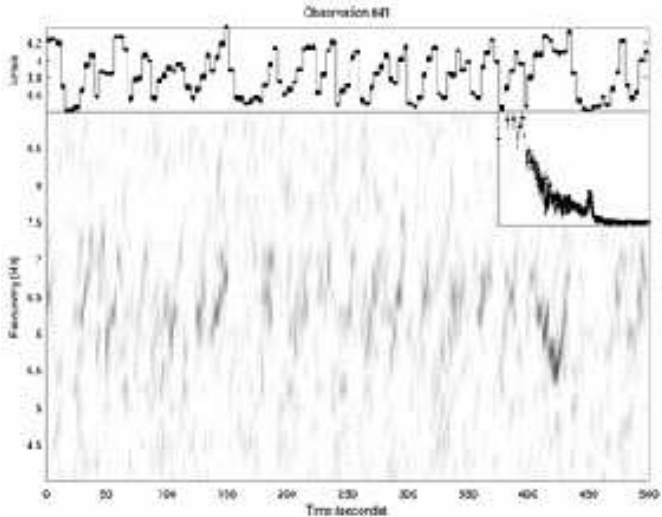}}
\caption{
Same as Fig. \ref{b5} for Obs. \#41.
} \label{b1}
\end{figure}

\begin{itemize}

\item {\it Observation \#41}: From Fig. \ref{b1}, it is evident that the 
        light curve (top panel) is characterized by oscillations between
        two flux levels. From the corresponding spectrogram (bottom panel)
        one can see that the QPO is present only during the high-flux
        intervals, where it is correlated with count rate. 
        From the spectrogram, we produced two average power spectra 
        by selecting 
        all power spectra corresponding to a count rate $<$3400 cts/s and 
        $>$3500 cts/s respectively (3 PCUs were on during this observation).
        In the power spectrum corresponding to the low count-rate 
        intervals (see Fig. \ref{b1}), no QPO
        is indeed visible: a fit with a zero-centered Lorentzian yields
        a characteristic frequency $\nu_b$=2.68$\pm$0.32 Hz and a fractional
        rms of $\sim$8\%.
        In the power spectrum corresponding to high count 
        rates, a narrow QPO is 
        visible. A Gaussian fit yields $\nu_q$=6.45$\pm$0.03 Hz
        and $\sigma$=0.52$\pm$0.03 Hz, with an integrated fractional rms
        of $\sim$5\%. A band-limited noise component is also seen: a Lorentzian
        fit gives $\nu_b$=5.74$\pm$0.44 Hz and a fractional
        rms of $\sim$6.5\%. Additional power is present at low frequencies
        ($<$1 Hz). The integrated 0.1--64 Hz fractional rms is similar
        between the low and high count rate power spectra.
        
\begin{figure}[h] \resizebox{\hsize}{!}{\includegraphics{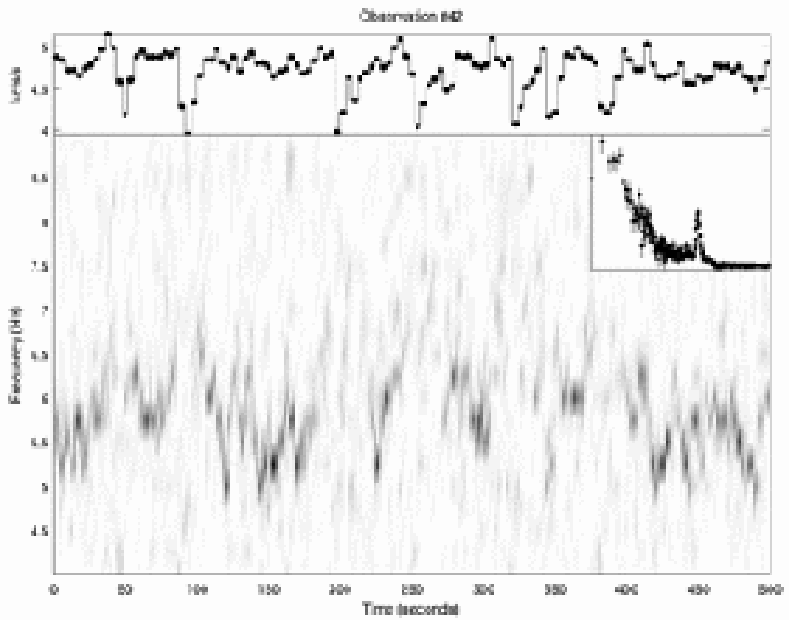}}
\caption{Same as Fig. \ref{b1} for Obs. \#42.
} \label{b2}
\end{figure}
        
\item {\it Observation \#42}: here the situation is slightly different. The
        QPO is only visible in the high-flux intervals, as for Obs. \#41,
        but the low-flux intervals look more like dips in the light curve
        as they are relatively few and sparse. Again, we extracted two
        average power spectra from the 4-s spectrogram (see Fig. \ref{b2}).
        During the dips (rate $<$4200 cts/s for 4 PCUs), no significant
        QPO is visible: a good fit is obtained with a broad Lorentzian
        component with $\nu_{max}$=4.23$\pm$0.39 Hz, plus an excess at
        low frequencies. Outside the dips, the narrow QPO appears at
        an average frequency of $\nu_q$=5.79$\pm$0.02 Hz (Gaussian fit) with
        $\sigma$=0.39$\pm$0.02 Hz and with an integrated fractional rms
        of $\sim$5.6\%. A subharmonic and a second harmonic components are
        also evident. The continuum noise can be approximated with a broad
        Lorentzian component ($\nu_{max}$=6.62$\pm$0.34 Hz), with as usual
        some excess at low frequencies.
        As for Obs. \#41, the integrated 0.1--64 Hz fractional rms is similar
        between the low and high count rate power spectra.
        
\begin{figure}[h] \resizebox{\hsize}{!}{\includegraphics{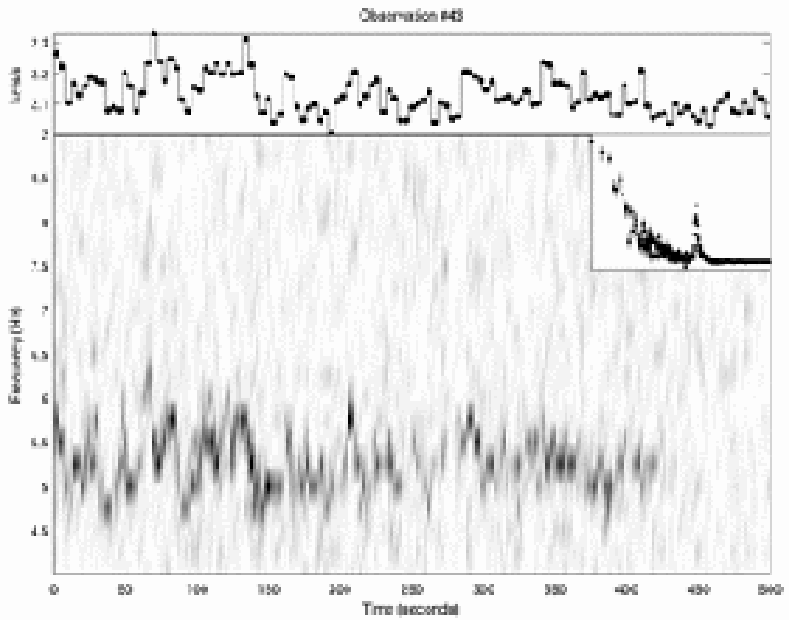}}
\caption{Same as Fig. \ref{b1} for Obs. \#43.
} \label{b3}
\end{figure}    

\item {\it Observation \#43}: from Fig. \ref{b3} the presence of a similar
narrow QPO peak is evident. The observation consists of three separate 
segments of $\sim$3400 s exposure each separated by 2200s-long gaps.
The QPO is present only in the first $\sim$1550 seconds of the second
interval, after which it disappears on a time scale of a few seconds
(see Fig. \ref{b3}). The QPO is clearly related to additional low-frequency
noise in the light curve (see Fig. \ref{b3}). Accumulating the 128s 
power spectrum
corresponding to the interval when the QPO is present (in this case the 4s
resolution is not needed), a clear peak is present. We fitted it with a
Gaussian with centroid frequency $\nu_q$=5.35$\pm$0.01 Hz and 
$\sigma$=0.38$\pm$0.01 Hz, for an integrated fractional rms of 7.7\%.
The continuum noise can be fitted with two zero-centered Lorentzians with
characteristic frequencies 0.02$\pm$0.01 Hz and 7.38$\pm$1.16 Hz and
fractional rms 3\% and 4.2\% respectively. The low-frequency component is
clearly the one responsible for the noise visible in Fig. \ref{b3}.
The average power spectrum 
outside the QPO interval is quite different. It can be fitted
with two zero-centered Lorentzians plus a broad QPO peak (see also
Nespoli et al. 2003). The QPO has a centroid frequency of 
$\nu_q$=6.71$\pm$0.14 Hz and a $\Delta$=3.27$\pm$0.46, for an integrated
fractional rms of 2.2\%. The two noise components have
 characteristic frequencies 0.05$\pm$0.01 Hz and 0.94$\pm$0.19 Hz and
fractional rms 1.6\% and 1\% respectively. Interestingly, there is a very
narrow excess over the best-fit model at 6.6 Hz, but an additional 
component is not required by the fits and nothing obvious is visible in 
the spectrogram.
Here the type-B QPO power spectrum has a 0.1--64 Hz integrated fractional rms
of 9.3\%, while for the broad QPO power spectrum 
it is 3.2\%. Since the hardness 
increases only slightly between the two intervals, notice that the 3.2\% value 
occupies the position of the `type A' group, suggesting that the broad
QPO is of type A (see below).

\begin{figure}[h] \resizebox{\hsize}{!}{\includegraphics{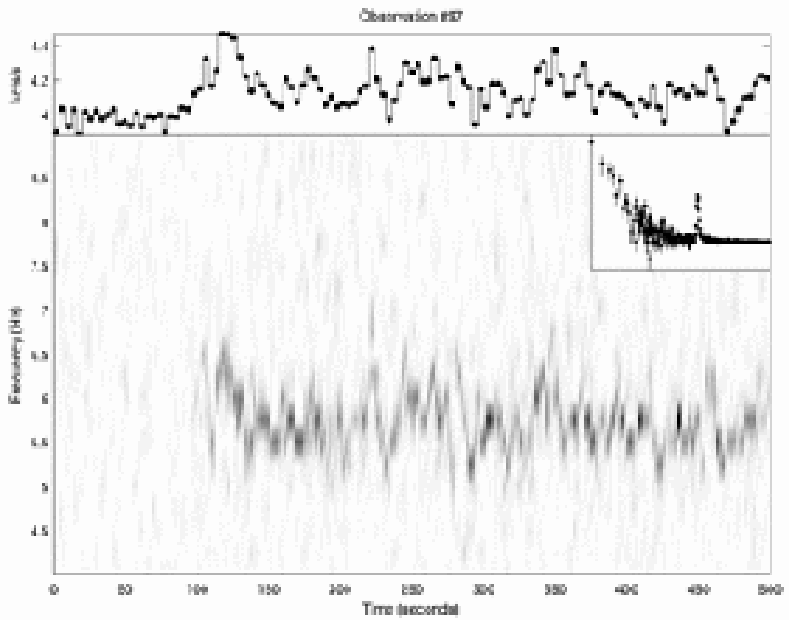}}
\caption{Same as Fig. \ref{b1} for Obs. \#97.
} \label{b4}
\end{figure}

\item {\it Observation \#97}: much later in the outburst, the sharp type-B
QPO appears again in this observation. The first part of the observation
$\sim$850s shows no evidence of the QPO, which appears suddenly together
with additional low-frequency noise (see Fig. \ref{b4}), with a variable
centroid frequency once again correlated with the count rate. All these
properties are consistent with those observed in previous observations,
despite the fact that this observation takes place almost four months 
after Obs. \#43 (see Tab. 1). We produced a spectrogram with time resolution 
16 seconds and averaged the power spectra 
corresponding to the presence and absence
of QPO. The QPO is fitted with a Lorentzian with 
$\nu_q$=5.71$\pm$0.02 Hz and a $\Delta$=0.44$\pm$0.03, for an integrated
fractional rms of 6.5\%. It is accompanied by a clear second harmonic, and
by a band-limited noise component with characteristic frequency
0.07$\pm$0.02 Hz and integrated fractional rms $\sim$6\%.
In the first part of the observation, when the QPO is not present, the 
average power spectrum is rather noisy: it can be fitted with  a broad 
($\Delta$=4.05$\pm$1.15 Hz) Lorentzian peak with a characteristic frequency
of 7.41$\pm$0.43 Hz and an integrated fractional rms of 2.1\%, plus a
band-limited noise with characteristic frequency 0.10$\pm$0.04 Hz and
fractional rms 1.4\%.
As for Obs. \#43, the integrated 0.1--64 Hz fractional rms is much lower
when the type-B QPO is not present: with a value of 3.0\%, it brings the
corresponding point in the `type A' group, although no significant QPO
is present.

\end{itemize}

\end{document}